\documentstyle[preprint,aps,epsf,tighten]{revtex}
\begin{document}
\draft
\preprint{
\vbox{\hbox{TPI-MINN-97/29-T}
      \hbox{UMN-TH-1613-97}
      \hbox{UND-HEP-98-BIG\hspace*{.2em}02} 
      \hbox{hep-ph/9805241}
      \hbox{(revised)}}}
\title{
Heavy flavor decays, OPE and duality in
two-dimensional 't~Hooft model
}
\author{
I. Bigi$^{\:a}$, M. Shifman$^{\:b}$, N. Uraltsev$^{\:a{\rm -}c}$, and
A. Vainshtein$^{\:b,d}$
}
\address{
$^a$Dept.of Physics, Univ. of Notre Dame du Lac, Notre Dame, IN 46556\\
$^b$Theoretical Physics Institute, Univ. of Minnesota,
Minneapolis, MN 55455\\
$^c$Petersburg Nuclear Physics Institute,
Gatchina, St.~Petersburg 188350, Russia
\footnote{Permanent address}
\\
$^d$Budker Institute of Nuclear Physics, Novosibirsk
630090, Russia
}
\maketitle
\thispagestyle{empty}
\setcounter{page}{0}
\begin{abstract}
The 't~Hooft model (two-dimensional QCD in the limit of large number
of colors) is used as a laboratory for exploring various aspects of
the heavy quark expansions in the nonleptonic and semileptonic decays
of heavy flavors. We perform a complete operator analysis and
construct the operator product expansion (OPE) up to terms 
${\cal  O}(1/m_Q^4)$, inclusively. The OPE-based predictions for the
inclusive widths are then confronted with the ``phenomenological"
results, obtained by summation of all open exclusive decay channels,
one by one. The summation is carried out analytically, by virtue of
the 't~Hooft equation. The two alternative expressions for the total
widths match. We comment on the recent claim in the literature of a
$1/m_Q$ correction to the total width which would be in clear conflict
with the OPE result.

The issue of  duality violations both in the simplified setting of
the 't~Hooft model and in actual QCD is discussed.  The amplitude of
oscillating terms is estimated.
\end{abstract}

\tableofcontents
\newpage

\section{Overview}
\label{sec:overview}

The development of heavy quark theory started in the 1980's, has
essentially been completed. While at the early stages the main
emphasis was placed on the symmetry aspects (the so-called heavy quark
symmetry), the present (mature) stage deals with dynamical aspects.  A
formalism based on Wilson's operator product expansion (OPE) \cite{1}
has been developed and applied to many cases of practical interest, in
particular to inclusive decays of heavy flavor hadrons.  The theory
of such decays is at a rather advanced stage now (see \cite{optical}
and references therein).  Calculations we could not even dream of
several years ago have become possible.

The decays of heavy flavor hadrons $H_Q$ are shaped by nonperturbative
dynamics.  While QCD at large distances is not yet solved,
considerable progress has been achieved in this problem. The width of
an inclusive transition $H_Q \to f$ is expressed through an OPE. The
nonperturbative effects are then parameterized through expectation
values of various local operators ${\cal O}_i$ built from the quark
and/or gluon fields.  Observable quantities, such as semileptonic and
nonleptonic widths of heavy hadrons $H_Q$, are then given by
\begin{equation}
\Gamma_{H_Q}= \frac{1}{M_{H_Q}}\sum_i {\rm Im}\, c_i(\mu)
\langle H_Q|{\cal
O}_i(\mu)| H_Q\rangle
\label{Gamma}
\end{equation}
where $c_i$ are the OPE coefficients, and $\mu$ stands for a
normalization point separating out soft contributions (which are
lumped into the matrix elements 
$\langle H_Q|{\cal O}_i(\mu)|H_Q\rangle$) 
from the hard ones (which belong to the coefficient
functions $c_i$).

There are many subtle and interrelated issues, both conceptual and
technical, associated with the operator product expansion in QCD:
\begin{enumerate}
\item 
Eq.~(\ref{Gamma}) represents an expansion in powers of $1/m_Q$
  with $m_Q$ being the $Q$ quark mass and the coefficients $c_i$
  scaling like $ (1/m_Q)^{d_i-2}$ for an operator with dimension
  $d_i$. In $\Gamma_{H_Q}$ there are two sources for contributions
  depending on powers of $1/m_Q$, namely higher-dimensional 
 {\em operators} and higher-order terms in the expansion of their 
  {\em expectation values}.  In addition every coefficient $c_i $ is a
  series in the running coupling $\alpha_s (m_Q) \propto 1/\log m_Q$,
  of which only a few terms are known for a given coefficient $c_i$.
  This immediately raises a grave concern: how can we retain terms
  suppressed by powers of $1/m_Q$ without a complete summation of the
  parametrically larger powers of $1/\log m_Q$ in the leading
  coefficient?
\item 
Although the normalization point $\mu$ conceptually represents a
  straightforward ``book-keeping" device for separating hard and soft
  contributions, it is technically difficult to actually carry out
  such a program since no user-friendly definition of what is soft and
  hard exists in QCD. So far, the vast majority of all discussions
  related to the introduction of $\mu$ are conducted in a hand-waving
  manner.

\item 
It is quite conceivable that there are {\em hard} 
  {\em non}perturbative contributions in the coefficient functions:
  $c_i^{nonpert} \sim (\Lambda_{\rm QCD}/m_Q)^\delta$ with $\delta $ being
  some positive number. The possible size of such contributions is
  essentially unknown.  A related problem is the convergence (or
  divergence) of the perturbative series for the coefficient functions
  $c_i$.

\item 
Truncating the series (\ref{Gamma}) at some finite order
  introduces an error estimated by  so-called exponential terms,
  which in Euclidean domain look as expressions of the type
  $\exp[-(m_Q/\Lambda_{\rm QCD})^k]$. In order to obtain $\Gamma_{H_Q}$ we
  analytically continue from the Euclidean domain, where the OPE is well
  defined and the coefficients $c_i$ are real, to the Minkowski domain
  where they acquire an imaginary part. Such analytic continuation is
  implicit in Eq.~(\ref{Gamma}) and is based on the assumption of
  smoothness. Under analytic continuation the exponential terms
  convert themselves into {\em oscillating} terms of the type ${\rm
    cos} [(m_Q/\Lambda_{\rm QCD})^k]$ \cite{Dike}; the expansion
  (\ref{Gamma}) does not account for them.  It can thus be understood
  on general grounds that duality violation is described -- or at
  least modeled -- by oscillating expressions.  To which degree those
  are suppressed by powers of $1/m_Q$ depends on details of the strong
  interactions and the specifics of the process.

\end{enumerate}

All these questions are circumvented in the so-called {\em practical
  version} of OPE \cite{2} routinely used so far in all instances when
there is need in numerical predictions.  This version is admittedly
approximate, however. The questions formulated above are legitimate;
they deserve to attract theorists' attention, and continue to cause
confusion in the literature. They have to be addressed also because
they are emerging as a major source of the uncertainties in
quantitative predictions; these problems have specifically been
suspected to underlie phenomenological difficulties encountered
recently, e.g.  a relatively short lifetime of beauty baryons and a
relatively small semileptonic branching ratio of beauty mesons.

We find it useful and instructive to study all these issues in models
that while retaining basic features of QCD -- most notably quark
confinement -- are simpler without being trivial and can be solved
dynamically. QCD defined in one time and one space dimension --
hereafter referred to as 1+1 QCD -- is especially suitable for this
purpose: with the Coulomb potential necessarily growing linearly in
two dimensions, quark confinement is built in.  Likewise the theory is
superrenormalizable, i.e. very simple in the ultraviolet domain. There
are no logarithmically divergent ``tails" in the Feynman graphs. As a
result, the book-keeping of OPE (separation of the hard and soft
parts) becomes simple, and all subtle aspects in the construction of
the OPE can be studied in a transparent environment.

In particular, the perturbative contributions in the coefficients
$c_i$ become an expansion in $g^2/m_Q^2$ (where $g$ is the gauge
coupling in 1+1 QCD). They are thus power-suppressed in the same way
as the higher-dimensional operators; the first problem formulated
above therefore does not arise here.  Without the logarithmic UV tails
the second problem becomes tractable. Concerning the third problem it
is easy to see that in 1+1 QCD nonperturbative corrections cannot
generate power suppressed terms in the coefficients $c_i$. For the
leading operator ${\bar Q}Q$ we will find its coefficient function to
all orders of perturbation theory (in the limit of $N_c \to \infty$),
demonstrating the convergence of the perturbative series.  At the same
time, the divergence of the condensate expansion in high orders will
become manifest indirectly, through the occurrence of oscillating
terms in $\Gamma_{H_Q}$, which appear with suppression factor
$(1/m_Q)^7$ in the case at hand.  Thus {\em all the four} problems
  formulated above will be answered!

We will perform our explicit calculations for 1+1 QCD in the limit of
a large number of colors $N_c$ -- the famous 't~Hooft model
\cite{H1,H2,LTLY}. For $N_c \to \infty$ only  planar diagrams
contribute in QCD; 1+1 QCD has the additional special feature that one
can choose a gauge such that there are  no gluon
self-interactions. Then only planar  ladder diagrams have to be
considered, and we have an exactly solvable theory in our hands. All
hadronic matrix elements of interest are therefore calculable.  This
enables us to describe every given transition in two complementary
ways: we can confront the OPE-based expression with a
``phenomenological" representation for the same process obtained by
saturating the rate by exclusive hadronic channels.

We want to take advantage of these unique features of the `t Hooft
model to illustrate all crucial elements of heavy quark theory and the
theory of inclusive heavy flavor decays in particular.  One should
keep in mind that heavy quark theory, as we know it now, is merely an
adaptation of the general OPE-based approach. Some of the questions to
be discussed below can therefore be actually formulated in a wider
setting.

The 't~Hooft model has been exploited as a theoretical laboratory for
testing various analytic QCD methods in applied problems before.
Heavy quark symmetry and heavy flavor decays were analyzed in 
Refs.~\cite{3,4,5}. The model was used recently for discussing general
aspects of OPE (convergence of the OPE series, exponential terms
violating duality, and so on)~\cite{6,7}.

In Ref.~\cite{5} heavy flavor inclusive widths were calculated 
numerically, by adding the exclusive channels one by one.  
It was
found that the inclusive width $\Gamma_{H_Q}$ approaches its
asymptotic (partonic) value, and the sum over the exclusive hadronic
states converges rapidly.  At the same time, small deviations from the
asymptotic value observed in the numerical analysis \cite{5} were
claimed to be a signal of $1/m_Q$ corrections in the total width, in
contradiction with the OPE-based result.

In this work we treat the very same problem, inclusive heavy
flavor decays in 1+1 QCD, {\em analytically}.  We first  develop
a technique perfectly parallel to that in four-dimensional QCD
\cite{optical}. It includes such elements as a complete operator
analysis and the construction of the transition operator.
Unlike four-dimensional QCD, the coefficient functions for the
leading  operator are exactly calculable (in the limit
$N_c\to\infty$).
Moreover, all relevant expectation values of the local operators
involved in the problem are calculable too. We get a
complete prediction through order $1/m_Q^4$.

Then we carry out a ``hadronic calculation"
of the same width, by saturating all open decay modes,
using the 't~Hooft equation \cite{H1}. By comparing the
phenomenological representation of the total width
with the OPE-based formula, we are able to identify, term-by-term,
the subsequent terms of the heavy quark expansion.
The situation actually turns out to be simpler than one could expect
{\em a priori}:
\begin{itemize}
\item In the $1/m_Q$ expansion for the inclusive width corrections of
  the order $(1/m_Q)^2$, $(1/m_Q)^3$ and $(1/m_Q)^4$ to the parton
  width come only from the leading operator $\bar Q Q$, i.e., from the
  expansions of its OPE coefficient $c_{\bar Q Q}$ and its expectation
  value $\langle H_Q |\bar Q Q |H_Q\rangle$. Operators of
  higher dimension contribute to the total width first at order
  $(1/m_Q)^5$.
\item 
The perturbative series in $g^2/m_Q^2$ for the OPE coefficient
  of the operator $\bar Q Q$ is completely defined by the one-loop
  renormalization of heavy quark mass. The result can be formulated in
  terms of the light-cone gauge formalism as the absence of
  renormalization.
\end{itemize}
These results are based on a general operator analysis. On the
``phenomenological" side, we use a sum rule, which is a consequence of
the 't~Hooft equation, to show that the total width is determined by a
quantity coinciding with the matrix element of the 
$c_{\bar Q Q}{\bar Q}Q$ term in the OPE expansion~(\ref{Gamma}), 
through order
$(1/m_Q)^4$. Thus, we observe a {\em perfect match} between the
  expression derived from the OPE and from adding up all relevant
  hadronic channels up to high order power corrections.

  After testing the validity of OPE, we exploit results obtained {\em
    en route} in order to discuss the issue of oscillating
  contributions related to the high-order tails in the OPE series that
  are factorially divergent.  Due to the simplicity of the model we
  can estimate them reliably.  A non-monotonous duality-violating
  component of the width for large $m_Q$ is suppressed by high power
  of $1/m_Q$ which we determined.  Implications of our analysis for
  real QCD are briefly discussed.

The remainder of the paper is organized as follows: after formulating
the problem in Sec.~\ref{sec:preliminaries} we construct the OPE and 
calculate the
coefficients in Sec.~\ref{sec:OPE}; after establishing the match between the
OPE-based result for the inclusive width and the sum rules for the
same width resulting from the 't~Hooft equation through order $(1/m_Q)^4$
in Sec.~\ref{sec:match}, we discuss an appearance of oscillating terms in the 
order
$(1/m_Q)^7$ and the duality violations they cause in 
Sec.~\ref{sec:violations}; in the
same section we discussed along similar lines a possible pattern of
the violation of the local duality for $\tau$ decays in 1+3
dimensions; in Sec.~\ref{sec:nonvanishing} we comment on 
the paper~\cite{5} and analyze
effects due to nonvanishing masses of light quarks; Sec.~\ref{sec:discussions} 
presents a
general discussion and conclusions.

\section{Preliminaries}
\label{sec:preliminaries}
We start by formulating the problem and introducing our notation
and conventions.

In two-dimensional QCD the Lagrangian looks superficially the same
as in four dimensions
\begin{equation}
{\cal L}_{1+1}=-\frac{1}{4g^2} \,
G_{\mu\nu}^a G_{\mu\nu}^a \,+\, \sum
\bar\psi_i
(i\not \!\!D -m_i)\psi_i \; , \; \;\;\;
i D_\mu=i\partial_\mu + A_\mu^a T^a\, ;
\label{11}
\end{equation}
$T^a$ denote generators of $SU(N_c)$ in the fundamental
representation, $G_{\mu\nu}^a$  the gluon field strength tensor
and $\psi_i$ the quark field ($i$ is a
flavor index) with a mass $m_i$; $g$ the gauge coupling constant. 

One
has to keep the following peculiarities in mind: $g$ carries dimension
of mass as does $\bar \psi \psi$. The field strength $G_{\mu\nu}^a$ on
the other hand has dimension $M^{2}$ in our normalization, just as in
four-dimensional QCD. With the theory being superrenormalizable no
(infinite) renormalization is needed; observables like the total width
$\Gamma_{H_Q}$ can be expressed in terms of the {\em bare} masses
$m_i$ and {\em bare} coupling $g$ appearing in the Lagrangian.
Anticipating the large $N_c$ limit we will use a parameter $\beta$
instead of $g$ where
\begin{equation}
\beta^2=\frac{g^2}{2\pi}\left( N_c -\frac{1}{N_c} \right)\;
\; \; {\rm with} \; \; \;
\lim_{N_c \to \infty} \beta ^2 = {\rm finite} \; .
\label{beta}
\end{equation}
This dimensionful quantity $\beta$, which -- in contrast to $m_i$ --
provides an  intrinsic mass unit for the 't~Hooft model, can be
seen as the analog of $\Lambda_{\rm QCD}$ of four-dimensional QCD.

We need at least two quarks denoted by $Q$ and $q$ with masses
$m_Q$ and $m_q$, respectively, to realize heavy flavor
transitions $Q \to q$. For quark masses we impose
\begin{equation}
m_Q - m_q \gg \beta\;,
\label{condm}
\end{equation}
where both $m_q \neq 0$ and $m_q=0$ are allowed for.  Condition
(\ref{condm}) guarantees that the inclusive methods of Ref.
\cite{optical} are applicable since it makes the energy release~
\footnote{It can hardly be overemphasized that it is the size of the
  energy release rather than of $m_Q$ that controls the reliability of
  the expansion in four dimensions as well.} in the weak decay large
relative to the intrinsic scale $\beta$.  We will actually employ the
dimensionless ratio $\beta/m_Q$ as our expansion parameter.

Next we need to introduce a flavor-changing weak interaction;
we  choose it to be of the current-current  form:
\begin{equation}
{\cal L}_{\rm weak}^V =-\frac{G}{\sqrt{2}}\,(\bar q \gamma_\mu Q)
\,
(\bar{\psi}_a\gamma^\mu \psi_b )\;.
\label{lweak}
\end{equation}
Here $G$ is an analog of the Fermi coupling constant; it is
dimensionless in two dimensions.  The fields $\psi_{a,b}$ can be
either the light quark or the lepton fields to describe nonleptonic or
semileptonic decays, respectively. In 1+1 dimensions the axial current
reduces to the vector one.  The most general current-current
interaction contains an additional term where the vector currents are
contracted via the antisymmetric $\epsilon_{\mu\nu}$ instead of
$g_{\mu\nu}$.  For the total width -- our main focus here -- such an
additional term is of no importance. The product of scalar densities,
on the other hand, is inequivalent to that of vector densities; we
will briefly discuss it, but mainly focus on the $V\times V$
interaction (\ref{lweak}).

For $N_c \to \infty$ factorization holds; i.e., the transition
amplitude can be written as the product of matrix elements of the
currents $\bar q \gamma_\mu Q$ and $\bar{\psi}_a\gamma^\mu \psi_b$. 
For the inclusive widths which are discussed below the property of 
factorization can be expressed as follows:
\begin{equation}
  \label{fact}
M_{H_Q}\Gamma_{H_Q}={\rm Im} \int {\rm d}^2 x \, i \,
\langle H_Q|\,T\left\{ {\cal L}_{\rm weak}(x)
{\cal L}_{\rm weak}^\dagger(0)\right\}|H_Q\rangle=
G^2\, \int {\rm d}^2 x \, {\rm Im}\Pi_{\mu\nu}(x)\,  
{\rm Im}T^{\mu\nu}(x)
\end{equation}
where $\Pi_{\mu\nu}(x)$ and $ T_{\mu\nu}(x)$ are defined as
\begin{equation}
  \label{Pi1}
  \Pi_{\mu\nu}(x)= i\,\langle 0|T \left \{ \bar{\psi}_a(x) \gamma_\mu \psi_b(x)
\, \bar{\psi}_b(0) \gamma_\nu \psi_a(0)\right\}|0\rangle\;,
\end{equation}
\begin{equation}
  \label{T1}
T^{\mu\nu}(x)= i\,\langle H_Q |T \left \{ \bar{q}(x) \gamma^\mu Q(x)\,
\bar{Q}(0) \gamma_\nu q(0)\right\}|H_Q\rangle\;.
\end{equation}
This factorization follows from the fact that at $N_c\to \infty$ there is 
no communication between $ \bar{\psi}_a \gamma_\mu \psi_b$ and 
$ \bar{q} \gamma^\mu Q$ currents: any gluon exchange brings in 
a suppression factor $1/N_c^2$.

The only difference between the semileptonic and nonleptonic widths
resides in $\Pi_{\mu\nu}(x)$, in the first case $\psi_a$ are leptonic fields
while in the second case they are the quark fields. 
At $m_\psi =0$ we get one and the same $\Pi_{\mu\nu}(x)$ (up to the overall
normalization factor $N_c$) as we will show shortly.
For
this reason at $m_\psi =0$ the distinction between the nonleptonic and 
semileptonic
cases is actually immaterial. At $m_\psi \neq 0$ quark and lepton 
polarization tensors $\Pi_{\mu\nu}(x)$ become different. This difference 
is proportional to powers of $m_\psi$. It will be discussed in 
Sec.~\ref{sec:nonvanishing}. For the time being we will treat $\psi$'s as 
massless leptons.

\begin{figure}[h]
\centerline{\epsfbox{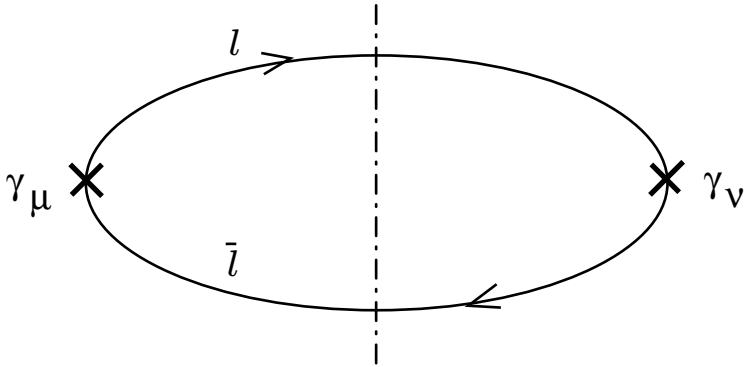}}
\caption{Polarization operator for lepton current}
\label{polariz}
\end{figure}

The one-loop graph determining $\Pi_{\mu\nu}$ is depicted in
Fig.~\ref{polariz}.
\firstfigfalse
For a massless fermion $\psi$ we get the well-known expression,
\begin{equation}
\Pi_{\mu\nu} (q)=\int {\rm d}^2 x \, e^{iqx} \Pi_{\mu\nu}(x)
=-\frac{1}{\pi }\, \left(\frac{q_\mu q_\nu }{q^2} - g_{\mu\nu}
\right)\, .
\label{piq}
\end{equation}
This expression obtained from a one-loop graph is known to be
exact. If $\psi$ is a lepton field, this statement is trivial.
If $\psi$ is the quark field 
 all gluon insertions inside the loop
automatically vanish  due to special properties of the
two-dimensional $\gamma$ matrices \footnote{Namely, one uses the
fact that $\gamma^\alpha \gamma^\mu\gamma_\alpha = 0$ and
 any odd number of $\gamma$ matrices reduces to one.}.
Thus, at $m_\psi = 0$  the only distinction
between $\psi_{a,b}$ being quark
rather than lepton fields is an overall factor $N_c$ on the
right-hand side of Eq.~(\ref{piq}).

A remarkable feature of Eq.~(\ref{piq}) is the occurrence of the pole
at $q^2=0$, which is specific for the vector interaction.  This means
that a pair of massless leptons produced by the vector current is
equivalent to one massless boson, whose coupling is proportional to
its momentum $q_\mu$. In the case of the quark fields, it is known
\cite{H2} from the early days of the 't~Hooft model that the vector
current $\bar{\psi}_a \gamma^\mu \psi_b$ produces from the vacuum only
one massless meson, the pion. This is readily seen by inspecting the
't~Hooft equation \cite{H1}.

For all computational purposes the vector current $\bar{\psi}_a
\gamma^\mu \psi_b$ in Eq.~(\ref{lweak}) can thus be substituted by
$\epsilon^{\mu\nu} \partial_\nu \phi/\sqrt{\pi}$ where $\phi$ denotes
a pseudoscalar massless noninteracting field,
\begin{equation}
{\tilde{\cal L}}_{\rm weak}^V =-\frac{G}{\sqrt{2\pi}}\;
\bar q \gamma_\mu Q \, \epsilon^{\mu\nu} \partial_\nu \phi \, .
\label{tildel}
\end{equation}
In other words, the problem is formulated as the inclusive decay
of the heavy quark $Q$ into a lighter quark $q$ plus a sterile  boson
$\phi$.
More exactly, we deal with the decays of  a $Q$ containing hadron
$H_Q$ into a $q$ containing final hadronic state $X_q$ plus $\phi$.

Let us pause here for two remarks. {\bf (i)} The fact that the
interaction vertex of the massless field $\phi$ involves
$\epsilon^{\mu\nu}$, see Eq.~(\ref{tildel}), is most obvious when
$\psi_{a,b}$ are quark fields.  For in this case $\phi$ is the pion,
as mentioned above, and the pion is coupled to the vector current
$\bar\psi_a\gamma_\mu\psi_b$ obviously through $\epsilon^{\mu\nu}$.
The case of the leptonic fields $\psi_{a,b}$ is indistinguishable;
therefore, the coupling of $\phi$ is the same.  {\bf (ii)} To keep the
analysis to be presented below as clean and transparent as possible we
want to be free of annihilation and Pauli interference contributions
to the total width (at least through ${\cal O}(1/m_Q^4)$).  This is
readily achieved by assuming throughout the paper that the spectator
light quark $q_{sp}$ in $H_Q$ is distinct from $q$.

\begin{figure}[h]
\centerline{\epsfbox{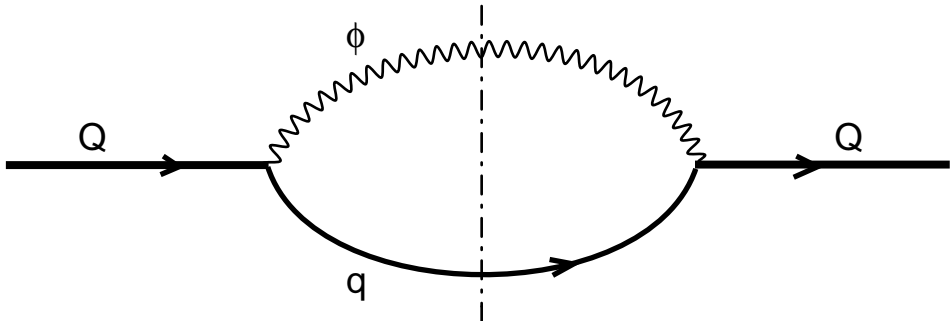}}
\caption{Transition operator in the leading order. The wavy line of
massless
$\phi$ field substitutes the propagation of lepton pair}
\label{transition}
\end{figure}

In the leading approximation the transition operator is determined by
the diagram of Fig.~\ref{transition}, where the wavy line corresponds to the
$\phi$ quantum. A straightforward calculation yields for the transition
operator
\begin{equation}
\hat{T}_0=c_{{\bar Q} Q}^0 \; {\bar Q} Q\,;~~~
2\,\mbox{Im}\, c_{{\bar Q} Q}^0  = \Gamma_Q = \frac{G^2
}{4\pi}\cdot
\frac{m_Q^2-m_q^2}{m_Q}\; ,
\label{cqqzero}
\end{equation}
where $\Gamma_Q$ is the decay width for a free quark Q as
evaluated
in the parton model.
This parton expression  will serve as reference in
analyzing the $(1/m_Q)^n$ corrections to the total width $\Gamma$.

\section{Operator product expansion for inclusive widths}
\label{sec:OPE}
\subsection{Catalogue of operators}
\label{sub:catalog}
The $1/m_Q$ expansion for inclusive widths of heavy flavor hadrons is
constructed from the Lorentz invariant weak transition operator
\cite{Vol}
\begin{equation}
 \hat{T} (Q\to Q) \;=\; \int {\rm d}^2x\; i T\left\{ {\cal L}_{\rm weak}(x)
{\cal L}_{\rm weak}^\dagger (0)\right\} = \sum c_i(\mu){\cal
O}_i(\mu)
\, .
\label{5}
\end{equation}
The local operators ${\cal O}_i$ are ordered according to their
dimensions. The leading one is $\bar{Q}Q$ with dimension 
$d_{\bar{Q}Q}=1$.  
Higher operators have dimensions $d_i >1$. By dimensional
counting the corresponding coefficients are proportional to
$(1/m_Q)^{(d_i -2)}$.  The ratio of the coefficients
$c_i/c_{\bar{Q}Q}$ is proportional to $m_Q^{-d_i + d_{\bar{Q}Q}}$.

The coefficients $c_i$ are determined in perturbation theory as a
series in $\beta^2/m_Q^2$. It is crucial that these coefficients are
saturated by the domain of virtual momenta $\sim m_Q$ and are infrared
stable by construction. (All infrared contributions reside in the
matrix elements of the operators ${\cal O}_i$.) At this point we
should mention a drastic distinction between four- and two-dimensional
QCD.  In four dimensions the expansion parameter for the coefficients
is the running coupling $\alpha_s (m_Q )$; nonperturbative
contributions to the coefficients coming from distances $\sim 1/m_Q$
could in principle show up in the form $\exp (-C/ \alpha_s (m_Q ))
\sim (\Lambda_{\rm QCD}/m_Q )^\delta$ where $\delta$ is some unknown positive
index, not necessarily integer.  In two-dimensional QCD such terms
cannot appear: an analog of the exponential term above would be
$\exp (- Cm_Q^2 /\beta^2 )$.

A note concerning the choice of the normalization point $\mu$: we will
imply that
\begin{equation}
m_Q\gg \mu \gg \beta\;.
\end{equation}
In this range there is no real dependence on $\mu$, so we will
suppress the argument $\mu$ both in the coefficient functions and
operators.
 
The coefficient functions are not the only source of a $m_Q$
dependence.  The matrix elements of the operators ${\cal O}_i$ contain
an implicit $m_Q$ dependence too. (We recall that in our formalism,
unlike HQET \cite{HQET}, the fields of which the operators ${\cal O}_i$ 
are built are the standard Heisenberg operators, rather than
asymptotic in $m_Q$.) In particular, for the leading operator 
$\bar Q Q$ we have the following relation:
\begin{equation}
\int{\rm d}^2x\; \bar Q Q \;=\;
\int{\rm d}^2x\; \left\{
\bar Q \gamma_0 Q \,+\, \bar Q
\frac{\pi_\mu \pi^\mu +\frac{i}{2} \sigma^{\mu \nu}G_{\mu
\nu}}{2m_Q^2} Q
\right\}\;,
\label{13}
\end{equation}
where $\pi_\mu=iD_\mu - g_{\mu 0}\,m_Q$ and the integration over $x$
allows us to omit terms which are total derivatives.

In the rest frame of the hadron $H_Q$ the expectation value of 
$\bar Q \gamma_0 Q$ counts the number of $Q$ quarks,
\begin{equation}
\frac{1}{2M_{H_Q}} \langle H_Q | \bar Q \gamma_0 Q |H_Q\rangle
=1\, .
\end{equation}
The factor ${1}/{2M_{H_Q}}$ will be present in all matrix elements; it
corresponds to a relativistic normalization of the states,
\begin{equation}
\langle H_Q (\vec{p}\, ')| H_Q(\vec{p})\rangle =
2 E_{H_Q}\delta (\vec{p}\, ' - \vec{p}
\, )\, .
\end{equation}

From relation~(\ref{13}) the matrix element of $\bar QQ$ is therefore
unity, up to a quadratic correction:
\begin{equation}
\frac{1}{2M_{H_Q}} \langle H_Q | \bar Q  Q |H_Q\rangle =
1 + {\cal  O}\left(\frac{1}{m_Q^{2}}
\right)\, .
\end{equation}
Moreover relation~(\ref{13}) provides an operator form for $1/m_Q^2$
corrections. They come from $\bar Q \pi_\mu \pi^\mu Q/2m_Q^2$ 
which equals ${\bar Q} D_1^2 Q/2m_Q^2$ up to $1/m_Q^4$ corrections.  
Notice, that the operator $\bar Q D_1^2 Q$ is Lorentz
noncovariant and cannot enter directly into the OPE for the total
width but, as we see, enters indirectly through the matrix element of
the operator $\bar QQ$.

Eq.~(\ref{13}) contains also the `chromomagnetic' operator and it
looks as this operator contributes to $1/m_Q^2$ corrections. It is
not, however, the case. Indeed, this operator can be rewritten as
follows
\begin{equation}
{\cal O}_G = \frac{i}{2}\, \bar Q \sigma^{\mu\nu}G_{\mu\nu} Q
= \frac{i}{2}\, \bar Q \gamma_5  \epsilon^{\mu\nu}G_{\mu\nu} Q 
= \frac{i}{4}\, {\bar Q }\gamma_5 \epsilon^{\mu\nu}G_{\mu\nu} 
(1-\gamma^0) Q +\frac{i}{4}\,
{\bar Q}(1-\gamma^0)\gamma_5 \epsilon^{\mu\nu}G_{\mu\nu} Q
\;,
\end{equation}
where the relation $\sigma^{\mu\nu}=\gamma_5 \epsilon^{\mu\nu}$ is
used.  Taking advantage of the non-relativistic equations of motion to
replace $(1-\gamma^0) Q $ by $(1/m_Q) \gamma^1 iD_1 Q$ we get (up to
total derivatives)
\begin{equation}
{\cal O}_G = -\frac{1}{2m_Q}\bar Q  (D^\mu G_{\mu \nu}) \gamma^\nu Q  +
{\cal O}\left(\frac{1}{m_Q^2} \right)
 =
\frac{g^2}{2m_Q}\bar Q \gamma_\mu t^a Q \sum_q\,
\bar q \gamma^\mu t^a q+{\cal O}\left(\frac{1}{m_Q^2} \right)
\, ,
\label{og4q}
\end{equation}
where $t^a$ stand for the generators of the color group $SU(N_c)$.
Thus, the operator ${\cal O}_G$ reduces to a four-fermion operator
${\cal O}_{4q}$ with coefficient $g^2/m_Q$.

The absence of operators with the gluon field strength tensor
$G_{\mu\nu}$ in the OPE is a specific feature of two-dimensional QCD.
The physical reason for the reducibility of the gluonic operators is
the absence of real gluons in two dimensions. A particular consequence
of Eq.~(\ref{og4q}) is that in Eq.~(\ref{13}) the chromomagnetic
operator generates $1/m_Q^3$ terms only.

Thus  we come to the following representation for the matrix element 
of the leading operator ${\bar Q} Q$,
\begin{equation}
\label{QQ}
\frac{\langle H_Q | \bar Q  Q |H_Q\rangle}{2M_{H_Q}} =
1 - \frac{1}{2m_Q^2}\, \frac{\langle H_Q | {\bar Q} (-D_1^2) Q
|H_Q\rangle}{2M_{H_Q}}  + \frac{g^2}{2m_Q^3}\, \frac{\langle H_Q | 
\bar Q \gamma_\mu t^a Q \sum_q\,\bar q \gamma^\mu t^a q
|H_Q\rangle}{2M_{H_Q}} + {\cal O}\left(\frac{1}{m_Q^4} \right)
\end{equation}

\begin{figure}[h]
\centerline{\epsfbox{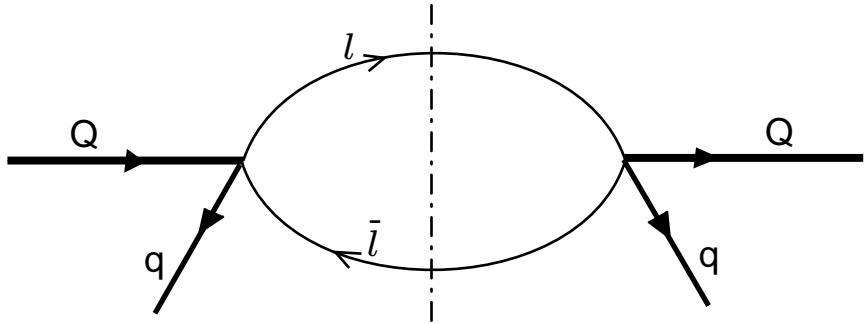}}
\caption{Four-fermion operators in the leading order}
\label{4q0}
\end{figure}

Let us proceed further with the operator analysis.  The first
subleading operator is the dimension-two four-fermion operator of the
type
\begin{equation}
\label{44q}
{\cal O}_{4q} = \bar Q \Gamma_1 Q \, \bar q  \Gamma_2 q\, ,
\end{equation}
where $\Gamma_{1,2}$ denote color and spinor matrices.  This is in
distinction with 1+3 QCD where the first subleading operator was 
$\bar Q \sigma^{\mu\nu}G_{\mu\nu} Q$.  On dimensional grounds the 
operator ${\cal O}_{4q}$, if present, could produce a {\em linear} 
$1/m_Q$ correction in the total decay width. To this end the
corresponding coefficient must arise in zeroth order in the coupling
$g^2$ (see the diagram in Fig.~\ref{4q0}).
For the $V\times V$ weak coupling of Lagrangian~(\ref{tildel})
this graph vanishes identically in the imaginary part,
Im~$c_{4q} =0$,
provided the leptons  are  massless, see discussion in 
Sec.~\ref{sec:preliminaries}.
(If the fields propagating in the loop are massive, the operator ${\cal
O}_{4q}$ appears. The corresponding modifications are considered in
Sec.~\ref{sec:nonvanishing}.)

 However, in the case of a scalar-scalar weak interaction of the form
\begin{equation}
{\cal L}_{\rm weak}^S =-\frac{G_S}{\sqrt{2}}\;
(\bar q  Q) \, (\bar{\psi}_a  \psi_b )
\label{ssl}
\end{equation}
the graph of Fig.~\ref{4q0} is not zero, and gives the following contribution
to the transition operator:
\begin{equation}
\mbox{Im}\, T_S = \frac{G_S^2}{4} \,  (\bar Qq)( \bar qQ) \, .
\label{imts}
\end{equation}
With this  contribution  the total width takes the form
(we put $m_q=0$ for simplicity):
\begin{equation}
\Gamma_{H_Q}^{S} = \frac{G_S^2 m_Q}{16\pi}\left(1 +
\frac{4\pi}{m_Q}\,
\frac{\langle H_Q|  (\bar Qq)( \bar qQ)|H_Q \rangle}{\langle H_Q|
 \bar QQ |H_Q
\rangle} \right) \, .
\label{1/m}
\end{equation}

Let us emphasize that, unlike four-dimensional QCD where
no operator can induce a $1/m_Q$ contribution to the
total width \cite{CGG,BUV}, this can happen in two dimensions.
The vanishing of Im~$c_{4q}$ for the $V\times V$ weak  Lagrangian in 
the leading order
is a specific dynamical feature of this particular Lorentz structure of
the weak interaction. 
Note at this point that
the argumentation presented in Ref.~\cite{CGG} was not sufficient 
to prove the absence of  $1/m_Q$ corrections in real QCD, see the
discussion in Sec.~\ref{sub:global}.

In Sec.~\ref{ssub:fourfermion}
we will show that Im~$c_{4q}$ vanishes not only in the leading order
in strong coupling
but also in  the order $g^2$. The first nonvanishing contribution to 
 Im~$c_{4q}$ comes in the order $g^4$ what leads to $1/m_Q^5$ corrections
to the width.

Next in the list comes the dimension-three 
operator containing six quark fields:
$$
(\bar Q\Gamma_1 Q )(\bar q_a\Gamma_2 q_b)
(q_c\Gamma_3 q_d)\;. 
$$ 
Along the same line of reasoning as for four-fermion operators we  show
that six-fermion operators appear only in the order $g^6$.  
It therefore contributes in order $1/m_Q^8$. For multi-fermion 
operators every extra ${\bar q} q$ pair brings in an extra $1/m_Q^3$ 
suppression.

In summary: to a quite high accuracy  the
 operator $\bar QQ$ is the only one to contribute.

\subsection{Calculating coefficients}
\label{sub:coefficients}
\subsubsection{Light-cone gauge and non-renormalization theorem}
\label{ssub:nonrenormalization}
The calculations are most conveniently done in the light-cone gauge.
This technology can be traced back to the pioneering work of 't~Hooft
\cite{H1}, and has been well studied in the literature.  In this
formalism the energy-momentum vector is described by
\begin{equation}
p^{\pm} = \frac{1}{\sqrt{2}} (E\pm p )\,  ,
\end{equation}
so that the mass-shell condition becomes $p^2 = 2p_+p_- = m^2$.

Let us write down the Lagrangian of the model in the light-cone
formalism:
\begin{equation}
{\cal L} = \sum_i \chi_i^{\dag} \left[ i\partial_ + -
\frac{m_i^2}{2i\partial_-} \right]
\chi_i + \frac{g^2}{2} \left( \sum_i \chi_i^{\dag} t^a \chi_i \right )
\frac{1}{\partial_-^2} \left( \sum_k \chi_k^{\dag} t^a \chi_k \right )
\label{lagr}
\end{equation}
In this formalism two-component quark $q_i$ fields are expressed via the
one-component fermionic fields $\chi_i$,
\begin{equation}
q_i = \frac{1}{2^{1/4}} \left(
\begin{array}{c}
\chi_i \\  \frac{m_i}{\sqrt{2} i\partial_-} \chi_i
\end{array}
\right)
\end{equation}
(in the basis where $\gamma_5=\gamma^0 \gamma^1$ is diagonal).  With
the gauge fixed by $A_-=0$, the $A_+$ component is expressed in terms
of the quark fields.

The weak interaction (\ref{tildel}) takes
the form:
\begin{equation}
{\tilde{\cal L}}_{\rm weak}^V =-\frac{G}{\sqrt{2\pi}}\;\left\{
\chi_q^{\dag} \chi_Q \,  \partial_+ \phi  -\frac{m_q m_Q}{2 }
\left[\frac{1}{i\partial_-}\chi_q\right]^{\dag}
\left[\frac{1}{i\partial_-}\chi_Q\right] \partial_- \phi \right\}\,.
\label{tilde2}
\end{equation}

A remarkable simplification occurs due to $\phi$ carrying light-like
momentum $q_\mu$: $q^2 = 2q_+q_- =0$.  We can satisfy this condition
by choosing the ``spatial" component of the momentum $q_- = 0$, i.e.,
$\partial_- \phi =0$. (This means that the $\phi$ quantum is a
left-mover.) Thus, on the $\phi$ ``mass shell'' the second term
(containing $\partial_-\phi $) in Eq.~(\ref{tilde2}) vanishes and the
weak $Qq\phi$ coupling takes a simple form,
\begin{equation}
{\tilde{\cal L}}_{\rm weak}^V =-\frac{G}{\sqrt{2\pi}}\;
\chi_q^{\dag} \chi_Q \,  \partial_+ \phi\;.
\label{tilde3}
\end{equation}

Here we come to a very important point. In the 't~Hooft model non-renormalization
theorem  for flavor non-diagonal currents at $q_-=0$  arises: {\em all} corrections
to the weak vertex at this point vanish. In the light-cone formalism the kinematical
point $q_-=0$ looks analogous to  the zero recoil point in four-dimensional heavy
quark theory.  The analogy is a superficial one, however~\footnote{
In  particular,  for  heavy-to-heavy transitions at $q_-=0$ there is no dominance
of the ground state production in the
't~Hooft model.
}. 
In 1+3 QCD flavor non-diagonal currents at the zero recoil are renormalized by
radiative and power corrections. 

\begin{figure}[h]
\centerline{\epsfbox{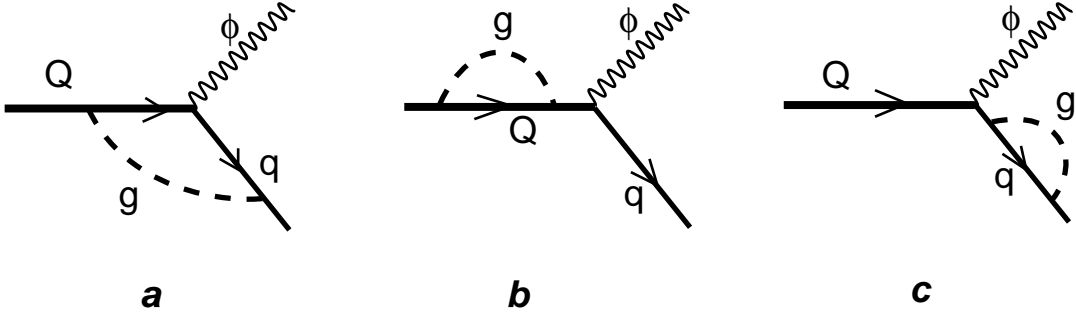}}
\caption{Radiative corrections to the weak coupling $Qq\phi$.
Dashed lines denote the gluons. }
\label{vertex}
\end{figure}

To prove the theorem let us consider the gluon corrections to the weak
coupling (\ref{tilde3}). To order $g^2$ the relevant graphs are
depicted in Fig.~\ref{vertex}.
The Feynman rules can be  read off from Eqs.~(\ref{lagr}),
(\ref{tilde3}).
In particular,
the weak $Qq\phi$ vertex is
\begin{equation}
V = -i \frac{G}{\sqrt{2\pi}}\, q_+ \;.
\label{V}
\end{equation}
The graph \ref{vertex}$a$ gives rise to following expression:
\begin{equation}
\Delta  V= V \cdot i\,\frac{\beta^2}{4\pi} \int {\rm d}^2 k \,
\frac{1}{k_-^2} \cdot \frac{1}{(p_q+k)_+ -
\frac{m_q^2 -i\epsilon}{2\,(p_q+k)_-}}
\cdot\frac{1}{(p_Q+k)_+ -\frac{m_Q^2 -i\epsilon}{2\,(p_Q+k)_-}}\;.
\label{dV}
\end{equation}
The integration over $k_+$ can easily be performed by the residue
method. It is clear then that the integration over $k_+$ produces a
nonzero result only in the case of opposite signs of $(p_q+k)_-$ and
$(p_Q+k)_-$ (the poles should be on the different sides of the
integration path).  For $q_- = (p_Q-p_q)_-=0$ these signs are
certainly the same, $(p_q+k)_-=(p_Q+k)_-$, and there is no correction
to the vertex.

To finish up with the $g^2$ correction to the weak coupling we need to
add graphs $b$ and $c$ of Fig.~\ref{vertex} containing the self-energies
$\Sigma_Q$ and $\Sigma_q$ of $Q$ and $q$ quarks.
\begin{equation}
\label{sigma}
\Sigma_Q(p_+,p_-) =-i\,\frac{\beta^2}{4\pi} \int {\rm d}^2 k \,
\frac{1}{k_-^2} \cdot \frac{1}{p_+ +k_+ -\frac{m_Q^2 -i\epsilon}{2\,
(p_-+k_-)}}\;.
\end{equation}
The integration contains a single pole only as a function of $k_+$.
Unlike the previously considered vertex correction, though, the
integration over $k_+$ gives a nonzero result because the integral
over the large semicircle in the complex plane of $k_+$ does not
vanish. The integration over $k_-$ requires an infrared regularization.
\footnote{OPE ensures that the dependence on the infrared regularization
disappears in the width as long as it is the same at all stages of the
calculation.}
Following 't~Hooft~\cite{H1} we define the integration in
Eq.~(\ref{sigma}) by putting a symmetric ultraviolet cutoff $K$ for
the $k_+$ integration, and a symmetric infrared cutoff $\lambda$ for
the $k_-$ integration,
\begin{equation}
|k_+|<K\;,~~~~~~~~~|k_-|>\lambda\;.
\end{equation}
Then at $|p_+|\ll K$  the result for $\Sigma$ is
\begin{equation}
\label{sigma1}
\Sigma_Q(p_+,p_-) =\left[ \frac{\beta^2}{2p_-}
-\frac{\beta^2}{2\lambda}
\,\epsilon(p_-)\right] \theta(|p_-|-\lambda)\;.
\end{equation}
The independence of $\Sigma$ on $p_+$ means that no $Z$ factor
appears.  The first term corresponds to a shift in the quark masses,
\begin{equation}
\label{masscor}
m_Q^2 \rightarrow m_Q^2 - \beta^2\:,~~~~m_q^2 \rightarrow m_q^2 -
\beta^2\;.
\end{equation}
The second term produces a (noncovariant) shift in the reference point
for the light-cone energy on mass shell,
\begin{equation}
\label{shift}
p_+= \frac{m_Q^2 - \beta^2}{2 p_-} + \frac{\beta^2}{2\lambda}\:,
~~~~~(p_->\lambda)\;.
\end{equation}
This shift produces no effect on the widths. One-loop radiative
corrections thus do not affect $Qq\phi$ transitions besides the mass
shift given by Eq.~(\ref{masscor}).

Moreover, it stays true for higher loops as well within the 't~Hooft
model.  For in the limit $N_c \rightarrow \infty$ there are no fermion
loop insertions into the gluon propagators. Then the higher loop
corrections to the vertex, as well as to the self-energy, vanish in
the way discussed above since the integration over $k_+$ yields zero.

Notice that the non-renormalization theorem we derive within
the 't~Hooft model is a
stronger statement than the one about zero recoil in four-dimensional
QCD where radiative and power
corrections break the non-renormalization of flavor non-diagonal
currents.
\subsubsection{The leading coefficient $c_{\bar QQ}$}
\label{ssub:leading}
Now it is simple to account for higher orders in the
 coefficient $c_{\bar QQ}$ of the leading operator ${\bar QQ}$.
To zeroth order in $g^2$ this coefficient was determined in 
Sec.~\ref{sec:preliminaries},
see Fig.~\ref{transition} and
Eq.~(\ref{cqqzero}). As just discussed higher loop corrections
merely shift the quark masses, Eq.~(\ref{masscor}), and therefore
we get the coefficient $c_{\bar QQ}$ in all orders,
\begin{equation}
  \label{corrqq}
  2\,\mbox{Im}\, c_{{\bar Q} Q}  = \frac{G^2 }{4\pi}\cdot
\frac{m_Q^2-m_q^2}{\sqrt{ m_Q^2 -\beta^2}}\; .
 \label{lnp}
\end{equation}
Combining this result with Eq.~(\ref{QQ}) and with the suppression of
the four-fermion operators, see Sec.~\ref{ssub:fourfermion}, we conclude that 
\begin{description}
\item[{\bf (i)}] there is no $1/m_Q$ corrections in the total width, much in
the same way as in actual QCD \cite{CGG,BUV}; 
\item[{\bf (ii)}] corrections 
$1/m_Q^2$, $1/m_Q^3$ and $1/m_Q^4$  to the total width 
are associated
exclusively with the operator $\bar Q Q$.
\end{description}
In Sec.~\ref{sec:match} 
we will prove a
stronger statement: irrespective of the explicit form of these
corrections, the hadronic saturation yields exactly the same result
for the total width as the contribution of $c_{\bar Q Q}\bar Q Q$ in
OPE.

The expression (\ref{lnp}) for $c_{\bar Q Q}$ refers to a low
normalization point, $\beta \ll \mu \ll m_Q$.  In order to calculate
the matrix element of $\bar Q Q$ over $H_Q$ we will need to express
$\bar Q Q$ in terms of $\chi_Q$. It goes without saying that the
operator $\bar Q Q$ must be taken at the same normalization point.
Then, the resulting series in $\beta^2/m_Q^2$ cancels against a
similar expansion coming from the operator ${\bar Q}Q$; there is no
$\beta$ dependence in the product $c_{{\bar Q} Q}\bar QQ$. This can be
seen by rewriting ${\bar Q} Q$ in terms of the unrenormalized
one-component field $\chi_Q$ and mass $m_Q$:
\begin{equation}
  \label{Qchi}
  {\bar Q} Q=\chi_Q^{\dag} \frac{m_Q}{i\partial_-}\chi_Q\;\,.
\end{equation}
In evolving down to $\mu$, higher orders lead to the substitution 
$m_Q \to \sqrt{m_Q^2 - \beta ^2}$ in this relation as well. With 
the quark mass substitution being the only effect of the radiative 
corrections we have
\begin{equation}
\label{Qchi1}
2\,\mbox{Im}\, c_{{\bar Q} Q} {\bar Q} Q = \frac{G^2 }{4\pi}
\,(m_Q^2-m_q^2)\, \chi_Q^{\dag} \frac{1}{i\partial_-}\chi_Q\, .
\label{msdop}
\end{equation}

The statement that the product $c_{{\bar Q} Q} {\bar Q} Q$ is
renormalization group invariant is trivial, of course.  A nontrivial
part of the result is encoded in Eq.~(\ref{msdop}), which is valid to
all orders in $g$. One could obtain this result by doing calculations
at $\mu \gg m_Q$ when the mass of the $Q$ quark coincides with its
``bare'' value $m_Q$, at $\mu = m_Q$, or at $\mu \ll m_Q$, when a
non-logarithmic evolution of the ${\bar QQ}$ operator and its
coefficient functions must be taken into account, the outcome is the
same, see Eq.~(\ref{msdop}).  To make contact with the 't~Hooft
equation (i.e. to calculate $\bar{Q}Q$ in terms of the 't~Hooft wave
function defined for bare quantities) we will need Eq.~(\ref{msdop})
at the ultraviolet cutoff.  Note that it can be conveniently rewritten
as
\begin{equation}
2\,\mbox{Im}\, c_{{\bar Q} Q} {\bar Q} Q
=
\Gamma_Q \,\chi_Q^{\dag} \frac{m_Q}{i\partial_-}\chi_Q
\, .
\label{msdopdop}
\end{equation}
In Sec.~\ref{sub:sumrules} we will find the matrix element of $\chi_Q^{\dag}
({m_Q}/{i\partial_-})\chi_Q$ and show that the corresponding
expression for the total width coincides with the one obtained
through the hadronic saturation.
\subsubsection{Four-fermion and multi-fermion operators}
\label{ssub:fourfermion}
\begin{figure}[h]
\epsfxsize 14cm
\centerline{\epsfbox{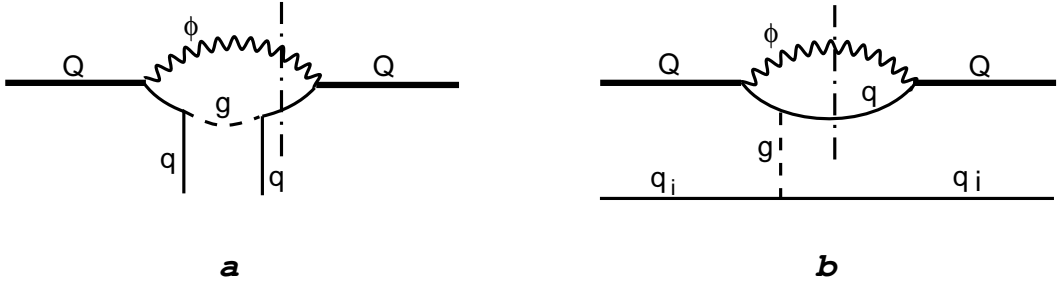}}
\caption{Four-fermion operators in $g^2$ order}
\label{4q2}
\end{figure}

As discussed above for the current-current weak interactions
four-fermion operators do not arise to zeroth order in the strong
coupling.
Diagrams generating four-fermion operators in the $g^2$ order are shown 
in Fig.~\ref{4q2}.
The light-cone gauge turns out to be again a convenient tool   
to use. Let us start for illustration with the simple
one-loop diagram of Fig.~\ref{transition}.  Feynman rules in the light-cone 
gauge, see Sec.~\ref{ssub:nonrenormalization}, lead to the following 
expression 
\begin{equation}
  \label{lc-trans}
  {\rm Im}\, i \int {\rm d}^2 k\, \delta \, (2 k_+ k_-)\, 
\frac{k_+^2}{(p_Q-k)_+ -
\frac{m_q^2 -i\epsilon}{2\,(p_Q-k)_-}}\;.
\end{equation}
Here $\delta \,(2 k_+ k_-)$ appears from the cut of the $\phi$ propagator,
the cutting of the $q$ quark propagator done by taking of imaginary part.
The integration over $k_-$ is immediate due to $k_-=0$ root of delta function
(the other root, $k_+=0$, gives the same contribution, so just a factor 2)
\begin{equation}
  \label{lc-trans1}
  {\rm Im}\, i \int {\rm d} k_+\, k_+\, 
\frac{1}{(p_Q-k)_+ -
\frac{m_q^2 -i\epsilon}{2\,(p_Q)_-}}
= \pi\left[ (p_Q)_+ - \frac{m_q^2 }{2\,(p_Q)_-}\right]
\;.
\end{equation}
It means that the term
\begin{equation}
  \label{lc-trans2}
  {\hat T}_0= {\rm const}\, 
\chi_Q^\dagger\left[ i\partial_+ - \frac{m_q^2 }{2\,i\partial_-}\right]\chi_Q
\end{equation}
appears in the transition operator.
In the zeroth order in the 't~Hooft coupling the equation of motion for
the $\chi_Q$ field is
\begin{equation}
\label{eqmotion}
\partial_+\chi_Q=m_Q^2/(2\,i\partial_-)\,\chi_Q 
\;.
\end{equation}
In the rest frame $i\partial_- \to p_-=m_Q/\sqrt{2}$,  and we reproduce
Eq.~(\ref{cqqzero}).

Let us now apply the same technique to the loop part in Fig.~\ref{4q2}$b$
in the limit of vanishing gluon momentum. Integrating over $k_-$ we get
\begin{equation}
  \label{lcloopb}
 {\rm Im}\, i \int {\rm d} k_+\, k_+\, 
\frac{1}{\left[(p_Q-k)_+ -
\frac{m_q^2 -i\epsilon}{2\,(p_Q)_-}\right]^2}=\pi
\;. 
\end{equation}
It produces the term 
\begin{equation}
  \label{lcloopb1}
  {\hat T}_1= {\rm const}\, 
\chi_Q^\dagger A_+ \chi_Q
\end{equation}
in the transition operator, note that the overall factor is the same as in 
Eq.~(\ref{lc-trans2}).
Summing up
${\hat T}_0$ and
${\hat T}_1$ results in the substitution of $i\partial_+$ by $i D_+=i\partial_+ + A_+$.
In this order the equation of motion~(\ref{eqmotion}) is also should be 
modified by the same substitution. Net result is that no change in OPE 
coefficients is produced by the loop in in Fig.~\ref{4q2}$b$.

Note that it is true not only for vanishing momentum of gluon field
but for any soft field as well, in other words, terms with derivatives of $A_\mu$
do not appear in the transition operator in one-loop order.
Note also that if extra gluons are emitted
 from the loop we get zero for the diagram. Indeed, it  increases the 
power in the integrand of Eq.~(\ref{lcloopb}) and the integral over the large 
semicirle in the complex plane of $k_+$  vanishes. It is the reason 
why diagrams of the type of  Fig.~\ref{4q2}$b$ give no rise to multi-fermion
operators.

To finish up with multi-fermion operators we need to account for diagrams of
the type given in  Fig.~\ref{4q2}$a$. The Feynman expression for this diagram
after integrating over $k_-$ is
\begin{equation}
  \label{lc-trans3}
  {\rm Im}\, i \int {\rm d} k_+\, k_+\, 
\frac{1}{\left[(p_Q-k)_+ -
\frac{m_q^2 -i\epsilon}{2\,(p_Q)_-}\right]^2}\; \frac{1}{[(p_Q-p_q)_-]^{2}}
=\frac{\pi}{[(p_Q-p_q)_-]^{2}}
\;.
\end{equation}
It is simple to check then that the diagram~\ref{4q2}$a$ cancel out 
against similar diagrams where the gluon exchange 
is between $q$ and $Q$ quarks. In case of 
 extra gluon insertions in one loop (relevant for six-fermion and higher 
dimension operators) we get a vanishing result right away.

\begin{figure}[h]
\epsfxsize 8 cm
\centerline{\epsfbox{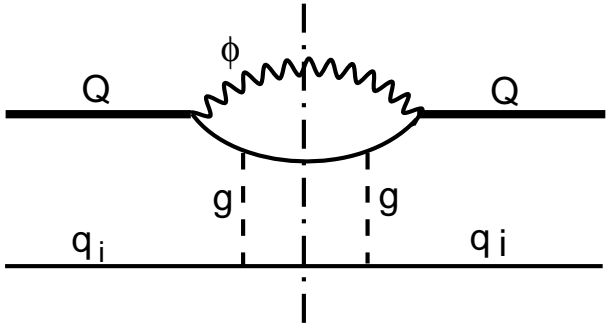}}
\caption{Four-fermion operators in $g^4$ order}
\label{4q4}
\end{figure}

Thus, we proved that four-fermion and multi-fermion operators do not arise 
at the level of one loop. They show up at level of the second loop,
see the
diagram~\ref{4q4} for four-fermion operator. The  dimensional counting 
reveals then that four-fermion operators give
$1/m_Q^5$ correction to the total width.

\subsection{OPE representation for inclusive width}
\label{sub:representation}

Putting everything together we  get
the OPE representation for the inclusive width:
\begin{equation}
\Gamma_{H_Q} =  \frac{G^2}{4\pi}\,
\frac{m_Q^2 - m_q^2}{\sqrt{m_Q^2 - \beta ^2}} \left[
\frac{\langle H_Q|\bar QQ |H_Q \rangle}{2M_{H_Q}} +
{\cal O}\left(\frac{1}{m_Q^5}\right) \right] =
\Gamma _Q \left[
\frac{\langle H_Q|\chi _Q^{\dagger}
\frac{m_Q}{i\partial _-}\chi _Q |H_Q \rangle}{2M_{H_Q}}
+ {\cal O}\left(\frac{1}{m_Q^5}\right)
\right]\, .
\label{GAMMADUAL}
\end{equation}
We have thus obtained a very simple result:
\begin{itemize}
\item
The partonic expression $\Gamma _Q$ represents
the asymptotic term for $m_Q \to \infty$.
\item
There is no $1/m_Q$ contribution in OPE 
as long as the weak interactions
are of the $V\times V$ type.
\item
Through order $1/m_Q^4$ only a single operator
contributes, $\bar QQ$.
\item
The leading correction $\sim {\cal O}(1/m_Q^2)$ enters
through the expectation value\\
$ (1/M_{H_Q})\langle H_Q |\chi_Q^\dagger ({m_Q}/{i\partial
_-})\chi _Q |H_Q \rangle$.
\end{itemize}
The second part of Eq.~(\ref{GAMMADUAL}) has been written in terms of
the light-cone operators to provide a way of rewriting the matrix
element in terms of the 't~Hooft wave function of the hadron $H_Q$.
The OPE result for the inclusive width can be recast in terms of the
sum over exclusive hadronic channels. This will be proven next.

\section{Match between OPE-based expressions and  hadronic saturation}
\label{sec:match}
\subsection{Exclusive widths via the  't~Hooft wave function}
\label{sub:exclusive}
With 1+1 QCD describing manifestly confining dynamics, its spectrum
consists of mesonic quark-antiquark bound states.  In $N_c \to \infty$
limit these mesons are stable in regard to strong decays. The masses
and the light-cone wave functions $\varphi (x)$ (with $x \,\epsilon\,
[0,1]$ having a meaning of a portion of momentum carried by the
quark) of these mesons can be determined as eigenfunctions and
eigenvalues of the 't~Hooft equation. In particular, the initial state
$H_Q=[Q{\bar q}_{sp}]$ is the ground state in the sector with the the
heavy quark $Q$ and the spectator antiquark ${\bar q}_{sp}$.  Its wave
function $\varphi_{H_Q}$ satisfies the following equation:
\begin{equation}
M^2_{H_Q}\varphi _{H_Q}(x) =
\left[
\frac{m_Q^2 - \beta ^2}{x} + \frac{m_{sp}^2 - \beta ^2}{1-x}
\right]
\varphi _{H_Q}(x) - \beta ^2 \int_0^1{\rm d}y \,\frac{\varphi
_{H_Q}(y)}{(y - x)^2}
\label{THOOFTIN}
\end{equation}
where $m_{sp}$ denotes the mass of the spectator antiquark and the
integral is understood in the principal value prescription.  The
solutions to the equation are singular at $x=0$ and $x=1$ where their
behavior is given by $x^{\gamma_0}$ and $(1-x)^{\gamma_1}$,
respectively, with $\gamma_{0,1}$ defined by the following conditions:
\begin{equation}
\frac{\pi \gamma_0}{{\rm tan}\pi \gamma_0} =
-\frac{m_Q^2-\beta^2}{\beta^2} \,\,, \;\;\;\;
\frac{\pi \gamma_1}{{\rm tan}\pi \gamma_1}=
-\frac{m_{sp}^2-\beta^2}{\beta^2} \,.
\end{equation}
The masses $M_n$ and wave functions $\varphi_n$ of final mesons 
$h_n =[q \bar q_{sp}]_n$ are defined by the same 't~Hooft equation with
$m_Q$ substituted by $m_q$:
\begin{equation}
M^2_n\varphi _n(x) =
\left[
\frac{m_q^2 - \beta ^2}{x} + \frac{m_{sp}^2 - \beta ^2}{1-x}
\right]
\varphi _n(x) - \beta ^2 \int_0^1{\rm d} y \,\frac{\varphi _n(y)}{(y - x)^2}
\label{THOOFTFI}
\end{equation}
The functions  $\varphi_n$  form a complete basis, i.e.,
\begin{equation}
\sum _n \varphi _n(x) \varphi _n(y) = \delta (x-y)
\label{ON}
\end{equation}

Let us find out now how an exclusive width $\Gamma_n$ of 
$H_Q\to h_n \phi$ decay is expressed via wave functions 
$\varphi_{H_Q}$, $\varphi_n$ of initial and final mesons,
\begin{equation}
\label{excl}
\Gamma _n = \frac{1}{2M_{H_Q}}\cdot
\frac{1}{(M^2_{H_Q} - M_n^2)}
\left| \langle h_n \phi|
{\frac{G}{\sqrt{2\pi}}\chi _q^{\dagger} \chi _Q \phi q_+}
|H_Q \rangle \right| ^2 =
\frac{G^2}{4\pi }
\frac{M_{H_Q}^2 - M_n^2}{M_{H_Q}}
\left[
\frac{\langle h_n|\chi _q^{\dagger} \chi _Q |H_Q \rangle}
{\sqrt{2}M_{H_Q}}\right] ^2
\end{equation}
where the factor $ 1/(M^2_{H_Q} - M_n^2)$ is the Lorentz invariant
phase space (LIPS) of two-particle final state and for the matrix
element we have used Eq.~(\ref{tilde3}) in the kinematics where the
$\phi $ momentum $q_\mu$ is
\begin{equation}
q_-= 0\, ,~~~~~~~~~~~~q_+ = \frac{1}{\sqrt{2}M_{H_Q}}\left( M^2_{H_Q}
- M_n^2 \right)\;.
\end{equation}
It is simple then to write the matrix element of $\chi _q^{\dagger}
\chi _Q $ in terms of the 't~Hooft wave functions,
\begin{equation}
\label{excl1}
\Gamma _n =  \frac{G^2}{4\pi }
\frac{M_{H_Q}^2 - M_n^2}{M_{H_Q}}
\left|
\int _0^1 {\rm d}x \varphi _n(x)\varphi _{H_Q}(x)
\right|^2 \;.
\end{equation}

\subsection{Sum rules}
\label{sub:sumrules}
Using the completeness
condition  (\ref{ON}) we can derive
sum rules for these partial widths by weighing them with
powers of
$M_{H_Q}^2 - M_n^2$. The first one is
\begin{equation}
\frac{4\pi M_{H_Q}}{G^2}
\sum _{n=0}^{\infty}\frac{\Gamma _n}{M_{H_Q}^2 - M_n^2}
 =\sum _{n=0}^{\infty}\left|
\int _0^1 {\rm d}x \varphi _n(x)\varphi _{H_Q}(x)
\right|^2=
\int _0^1 {\rm d} x \varphi _{H_Q}^2(x) = 1\;.
\label{SR1}
\end{equation}
Note that the sum runs over {\em all} states $h_n$ including those
unaccessible in the real decays of $H_Q$, i.e. with masses
$M_n>M_{H_Q}$. These transitions are still measurable by the
process of
inelastic lepton scattering off $H_Q$ meson. This sum rule is an
analog
of first Bjorken sum rule and was discussed in~\cite{Burkardt}.

To get next sum rules let us
multiply Eq.~(\ref{THOOFTIN}) by $\varphi _n(x)$ and
Eq.~(\ref{THOOFTFI}) by $\varphi _{H_Q}(x)$, respectively.
After integrating over $x$ and subtracting we find:
\begin{equation}
(M^2_{H_Q} - M^2_n)
\int_0^1 {\rm d} x \,\varphi _n(x) \varphi _{H_Q}(x) =
(m^2_Q - m^2_q)
\int_0^1 \frac{{\rm d}x}{x} \,\varphi _n(x) \varphi _{H_Q}(x)\;.
\label{HADQUA}
\end{equation}
Two more sum rules then arise:
\begin{equation}
\frac{4\pi M_{H_Q}}{G^2}
\sum _{n=0}^{\infty}
\Gamma _n =
(m_Q^2 - m_q^2) \int _0^1 \frac{{\rm d}x}{x} \,\varphi _{H_Q}^2(x)\;,
\label{SR2}
\end{equation}
\begin{equation}
\frac{4\pi M_{H_Q}}{G^2}
\sum _{n=0}^{\infty}
\Gamma _n (M_{H_Q}^2 - M_n^2)
=
(m_Q^2 - m_q^2)^2 \int _0^1 \frac{{\rm d}x}{x^2} \, \varphi _{H_Q}^2(x)\;.
\label{SR3}
\end{equation}
The second and third sum rules differ from the first one in two
aspects: they depend on the quark masses $m_Q$ and $m_q$ explicitly
and the integral over the wave function is not fixed by a
normalization condition; it can, however be calculated in the 't~Hooft
model.  One should note that while the integrand $\varphi
_{H_Q}^2(x)/x^2 \sim x^{-\beta ^2/m_Q^2}$, is singular at $x=0$, it is
still integrable, since $\beta^2/m_Q^2 \ll 1$.  Note also that
expanding the second sum rule~(\ref{SR2}) in $1/m_Q$ to the linear
order we reproduce the corresponding sum rule of Ref.~\cite{Burkardt}.

The sum rules above provide us with detailed information
on the saturation of the sums over the final hadronic state.
The quantity
\begin{equation}
  \label{proba}
  w_n= 
\frac{4\pi M_{H_Q}}{G^2}
\frac{\Gamma _n}{M_{H_Q}^2 - M_n^2}
\end{equation}
can be interpreted as a normalized probability of producing the state
$n$. Indeed $\sum w_n =1$ according to the first sum rule~(\ref{SR1}).
Then the sum rule~(\ref{SR2}) implies
\begin{equation}
  \label{average}
  \left\langle M_{H_Q}^2 - M_n^2 \right\rangle = 
(m_Q^2 - m_q^2) \int _0^1 \frac{{\rm d}x}{x} \, \varphi _{H_Q}^2(x)\;.
\end{equation}
If both masses $m_{sp}$ and $m_q$ are smaller or of order of $\beta$
we conclude that
\begin{equation}
  \label{satur}
\left\langle M_n^2 \right\rangle =\left\langle \frac{1}{x}-1 \right\rangle m_Q^2 +
{\cal O}(\beta^2) \sim \beta m_Q\;.
\end{equation}
Here we have anticipated the result for $\langle 1/x \rangle$ from
Eq.~(\ref{expan}) in the next subsection, in conjunction with 
Eq.~(\ref{Mh}). The reason for $\langle M_n^2 \rangle \sim \beta m_Q$ is
clear on the physical grounds: in the partonic approximation the final state 
is formed by the quark $q$ with the momentum $m_Q/2$ and by the
spectator antiquark ${\bar q}_sp$ with the momentum of order $\beta$. 

The sum rule~(\ref{SR3}) (after subtracting the square of the second 
sum rule~(\ref{SR2})) determines the dispersion:
\begin{equation}
  \label{disp}
\left\langle M_n^4 \right\rangle - \left\langle M_n^2 \right\rangle^2 = 
\left \langle 
\frac{1}{x^2}\right\rangle  -\left\langle \frac{1}{x}\right\rangle^2
\sim 
\left\langle M_n^2 \right\rangle^2
\;.
\end{equation}

What about higher moments? It is not difficult to see that the next
one, $\sum _{n=0}^{\infty} \Gamma _n (M_{H_Q}^2 - M_n^2)^2$, is a
divergent sum because $\varphi _{H_Q}^2(x)/x^3$ would no longer be
integrable. It defines the asymptotics of $\Gamma_n$ 
at large $n$,
\begin{equation}
\label{largen}
\Gamma _n \propto \frac{1}{M_n^6}\,.
\end{equation}
Let us recall~\cite{H1} that $M_n^2=\pi^2 \beta^2 n$ for high excitations, 
$n\gg 1$. We will see in the next subsection that the 
asymptotics~(\ref{largen}) matches the contribution of the four-fermion 
operators in OPE.

\subsection{Matching}
\label{sub:matching}
Armed with exact sum rules~(\ref{SR1}),~(\ref{SR2}),~(\ref{SR3}) we
are well prepared to verify a perfect match between the OPE result and
the expression obtained by summing over hadronic final states. From
the second sum rule~(\ref{SR2}) we have for the total semileptonic
width $\Gamma_{H_Q}$ the following relation
\begin{equation}
\label{width1}
\Gamma_{H_Q}= \frac{G^2}{4\pi}\cdot \frac{m_Q^2 -
m_q^2}{M_{H_Q}}
\int _0^1 \frac{{\rm d}x}{x}
\,\varphi _{H_Q}^2(x) - \sum _{M_n> M_{H_Q}} \Gamma _n \;.
\end{equation}
The second term is
actually  positive ($\Gamma_n$ defined by Eq.~(\ref{excl}) is
negative at $M_n>M_{H_Q}$). Using Eq.~(\ref{largen}) its size can be
estimated
\begin{equation}
- \sum _{M_n> M_{H_Q}} \Gamma _n \propto \frac{1}{m_Q^4}
\end{equation}

Thus we have derived from the 't~Hooft equation  the following result for
the inclusive width
\begin{equation}
\label{ghgq}
\Gamma_{H_Q}= \Gamma_Q \left[ \frac{m_Q}{M_{H_Q}}\int _0^1
\frac{{\rm d}x}{x}
\,\varphi _{H_Q}^2(x) + {\cal O}\left(\frac{1}{m_Q^5}\right) \right]\, .
\end{equation}
This expression coincides with the OPE result of Eq.~(\ref{GAMMADUAL})
as seen by rewriting the matrix
element in Eq.~(\ref{GAMMADUAL}) in terms of the ground state wave
function $\varphi_{H_Q}(x)$,
\begin{equation}\
\label{m.e.}
\frac{\langle H_Q|\chi _Q^{\dagger}
\frac{m_Q}{i\partial _-}\chi _Q | H_Q \rangle}{2M_{H_Q}}=
\frac{m_Q}{M_{H_Q}}\int _0^1 \frac{{\rm d}x}{x}
\,\varphi _{H_Q}^2(x)
\end{equation}
where, besides normalization factors, we have used also the
substitution $i\partial_- \to x M_{H_Q}/\sqrt{2}$.

This  completes the proof of the perfect matching between OPE and the
hadronic saturation through the order $1/m_Q^4$.  Let us stress that
the matrix element~(\ref{m.e.}) given by the integral over the ground
state wave function is implicitly $m_Q$ dependent, its leading term is
1 followed by  $1/m_Q^2$ and higher terms (see the discussion in 
Sec.~\ref{sub:catalog}).  
In the 't~Hooft model one can, of course, evaluate the matrix
element explicitly although it is not relevant to our main objective
-- probing the quark-hadron duality.

The absence of $1/m_Q$ corrections was demonstrated in 
Sec.~\ref{sub:catalog} by
operator methods. Let us show now that the same statement can be
derived from the 't~Hooft equation as well. To this end we use the
approach of Ref.~\cite{Burkardt} to the heavy quark limit,
generalizing it to include $1/m_Q^2$ corrections. Instead
of $x$, the appropriate variable for the large mass limit is
\begin{equation}
t=(1-x) m_Q\;,\;\;\;\; \varphi (x) = \sqrt{m_Q} \, \phi (t)
\end{equation}
An expansion of the 't~Hooft equation in $1/m_Q$ 
(after a substitution of the new variable) and the virial theorem
lead to the following relation for the meson mass,
\begin{equation}
\label{Mh}
\frac{M_{H_Q}^2}{m_Q^2}=1   + 2\frac{\langle t \rangle}{m_Q} -
\frac{\beta^2}{m_Q^2}
 + {\cal O}\left(\frac{1}{m_Q^3}\right)\;,
\end{equation}
where averaging is over the $H_Q$ meson wave function $ \phi (t)$. As
compared with Ref.~\cite{Burkardt1} we have added a $\beta^2/m_Q^2$
term.  Its origin is simple: it accounts for the renormalization of
the heavy quark mass.  Note that $\langle t\rangle \sim \beta$
provided $m_{sp}\lesssim \beta$.

We also need a similar expansion for the integral entering
Eq.~(\ref{ghgq}),
\begin{equation}
\label{expan}
\int _0^1 \frac{{\rm d}x}{x} \,\varphi _{H_Q}^2(x) = 1 + 
\frac{\langle t \rangle}{m_Q}
+ \frac{\left\langle t^2 \right\rangle}{m_Q^2} + 
{\cal O}\left(\frac{1}{m_Q^3}\right)\;.
\end{equation}
Substituting Eqs.~(\ref{Mh}, \ref{expan}) in Eq.~(\ref{ghgq}) 
we have,
\begin{equation}
\label{expansion}
\frac{\Gamma_{H_Q}}{\Gamma_Q}= 1+ \frac{1}{2m_Q^2} \left(
\beta^2 - \left\langle t^2 \right\rangle +\langle t \rangle^2 \right) + 
{\cal O}\left(\frac{1}{m_Q^3}\right)\;.
\end{equation}
This expansion should be compared with the operator
representation~(\ref{QQ}).  The $1/m_Q$ expansion of 
$c_{{\bar Q} Q}{\bar Q} Q$ produces the same $1/m_Q^2$ term; 
the part $\propto \beta^2$ comes from the expansion of 
$c_{{\bar Q} Q}$, see Eq.~(\ref{corrqq}).

Note that without the $\beta^2$ term the correction to 1 is negative,
i.e.
\begin{equation}
\frac{\Gamma_{H_Q}}{\Gamma_Q}-1< \frac{\beta^2}{2m_Q^2}\;.
\end{equation}

One more comment about $1/m_Q^5$ terms. In the OPE approach they are
due to the four-fermion operators generated by the graph in Fig.~\ref{4q4}.
Although the corresponding OPE coefficients are not calculated, the
consideration above shows that the contribution of the four-fermion
operators is dual to the sum of $\Gamma_n$ with $M_n> M_{H_Q}$ for
final state mesons, i.e. channels kinematically inaccessible in the
decay.

\section{Violations of duality}
\label{sec:violations}
\subsection{Global and local duality}
\label{sub:global}
Having established a perfect match between the OPE prediction for the
total width and the result of the saturation by exclusive decay modes,
through ${\cal O} (1/m_Q^4)$, we must now turn to the issue of where
the OPE-based prediction is supposed to fail.  The failure usually
goes under the name of ``duality violations", a topic under intense
scrutiny in the current literature.  The definition of what duality
violation is varies from publication to publication. Quite often, the
researchers in the field stick to a vague notion of deviations ``of
certain rates for processes involving hadrons from the underlying
partonic rates".  This is, for instance, the convention of
Ref.~\cite{5} where duality is understood as the coincidence with the
parton-model prediction.  If so, any nonperturbative contribution to
the given rate would be interpreted as a ``duality violation", which
does not make much sense to us.

We must precisely define what is meant by duality and its violations.
Assume that a certain process is amenable to calculations within OPE.
This means that an appropriate Euclidean quantity can be chosen, and
the OPE series can be constructed.  This series presents the quantity
of interest as an expansion in an inverse large parameter, e.g.
$1/Q^2$ or $1/E$.  The very same quantity can be expressed as a
dispersion integral over the imaginary part defined in Minkowski
space.  In $e^+e^-$ annihilation the imaginary part coincides with
$R(e^+e^-)$, in the transition amplitudes for heavy flavors the
imaginary part reduces to semileptonic spectral densities, etc.  

In
order to treat the nonleptonic decays in the same vein one can
introduce a spurion in the weak vertex, carrying a momentum $q$.
In other words, let us substitute the weak lagrangian by
\begin{equation}
  \label{spurion}
  {\cal L}_{\rm weak}(x) \to S(x){\cal L}_{\rm weak}(x)
\end{equation}
where $S(x)$ is a spurion field. Now let us consider the forward amplitude
${\cal A}(s)$ of the process
\begin{equation}
  \label{forward}
  S(q) + H_Q (p) \to \mbox{light hadrons} \to  S(q) + H_Q (p)
\end{equation}
as a function of $s=(p+q)^2$. This variable $s$ plays the same role 
as $s=q^2$ in $e^+e^-$ annihilation. The total nonleptonic  width
is given by ${\rm Im}\,{\cal A}$ at $s=M_{H_Q}^2$. We are free to consider
 ${\rm Im}\,{\cal A}(s)$ in the complex $s$ plane where it has two cuts:
at $s>0$ and at $s<-2M_{H_Q}^2+2q^2$ (the second cut is due to $u$ channel).
Choosing a reference point $s_0$ far away from both cuts but closer to the 
first one we can express ${\cal A}(s_0)$ as a dispersion integral over
the discontinuity across the cuts. On the other hand at the very same point
$s_0$ one can apply the operator product expansion for calculating 
${\cal A}(s_0)$ in terms of matrix elements of local operators,
$\langle H_Q |{\cal O}_i|H_Q\rangle$. This gives sum rules which allow us
to determine  ${\rm Im}\,{\cal A}(s)$ at large $s$, in particular,
at $s=M_{H_Q}^2$. This is fully analogous to what one does in $e^+e^-$
annihilation for $R(s)$. In both cases smoothness is assumed (of course,
in in $e^+e^-$ annihilation all positive values of $s$ are accessible
and one can check this assumption while in the case of nonleptonic
decays $M_{H_Q}$ is fixed).  Needless to say that the total semileptonic 
widths can be treated along the same lines, the only difference is 
the presence of leptons in the intermediate states.

In the semileptonic decays OPE allows to predict, additionally,
various distributions in the lepton momenta. This is, probably, the reason
why it is usually claimed that the status of duality is more solid
in the semileptonic decays. To show that it is not the case let us consider
the semileptonic decays with the light quark in the final state.
The OPE-based predictions for the spectral distributions are valid
almost everywhere; they fail only in the end-point domain~\cite{CGG}.
For this reason the total semileptonic widths cannot be obtained by 
integrating over the spectrum if we want a prediction which includes 
the linear in $1/m_Q$ corrections, it is why the argumentation in 
Ref.~\cite{CGG} was not sufficient to prove the absence of  
$1/m_Q$ corrections.  Nevertheless, the statement of absence
of such corrections in the total semileptonic widths can certainly 
be justified by the procedure described above for the nonleptonic widths.
Thus, we see that the theoretical status of all these processes --
$e^+e^-$ annihilation, semileptonic and nonleptonic decays of heavy flavors --
is basically the same.

By performing an appropriate expansion of the dispersion integral we
obtain sum rules relating certain moments of the imaginary part of
transition amplitude to matrix elements of consecutive terms in the
OPE series constructed in the Euclidean domain.  The predictions
obtained in this way will be referred to as {\em global duality}.  Taken
at their face value, they are exact, to the extent we can calculate
the coefficient functions and matrix elements of the operators
involved in OPE.  No additional assumptions are made. The predictions
obtained in this way are consequences of fundamental QCD. Therefore,
it does not make any sense to speak about violations of the global
duality.  One can only speak of the precision of calculation of the
coefficient functions and determination of the matrix elements.

Unfortunately the term {\em global duality} is often used in a loose
and ambiguous sense. It is applied indiscriminantly to integrals over
the spectral densities with the weight functions chosen {\em ad hoc}.
Our definition is narrower: it refers only to those specific integrals
which emerge from the dispersion representation.

The notion of {\em local duality} on the other hand requires further
assumptions.  Assume that we want to predict imaginary parts (spectral
densities) point by point, at large energies (or $q^2$). {\em If} one
assumes that the spectral densities at the given energy are smooth,
then from the moment integrals we can certainly predict the densities
themselves. This amounts to an analytic continuation of the OPE series
(truncated in a certain way), term by term from the Euclidean to
Minkowski domain, with the subsequent calculation of the imaginary
parts of each individual term in the series.  The prediction obtained in
this way is evidently a smooth function of the parameters. We then
compare this prediction with the quantity measured in terms of
hadronic contributions.  The difference between the OPE-based smooth
result and the experimental hadronic measurement is referred to as the
{\em duality violation} meaning the violation of {\em local duality}.

By its nature the OPE results are series in powers of
$\Lambda_{\rm QCD}/E$ and do not account for terms like $\exp[-
(E/\Lambda_{\rm QCD})^k]$ (in the Euclidean domain). Although such terms are
due to large distances, a signal of their appearance could show up in
the short distance OPE series in the form of a factorial divergence of
the series in higher dimensions. The situation is reminiscent of that
in the perturbative expansion. The divergent $\alpha_s$ series (e.g.
due to infrared renormalons) give rise to terms $\exp(-C/\alpha_s)$
although such terms can appear even in the absence of renormalons (for
instance, as the quark condensate).

In other words, the OPE construction accounts properly for short
distance singularities while the exponential terms are due to large
distances being nonsingular at short distances.  Thus, the duality
violation is something we do not see in the (truncated) OPE series.
The duality-violating terms are exponential in the Euclidean domain
and oscillating (like $\sin[ (E/\Lambda_{\rm QCD})^k]$) in the Minkowski
domain.

From this standpoint there is no distinction between, say, the total
$e^+e^-$ annihilation cross section or the semileptonic rates of heavy
flavors, on the one hand, and the nonleptonic rates of heavy flavors,
on the other.  Sometimes it is claimed that the former processes are
``pure" while the latter are ``impure"; it is even asserted that
``duality follows from OPE in the first case while it has no
theoretical justification in the second case".  We assert that a
duality violating exponential/oscillating component, associated with
the neglected tails of OPE, is present in all processes, and the only
physically meaningful question is its magnitude, as a function of
large parameters (e.g. a momentum transfer or $m_Q$) and specific
details of the process under consideration.

\subsection{Oscillating terms in 't~Hooft model}
\label{sub:oscillating}
The appearance of duality violations in the form of oscillating terms
is evident in the 't~Hooft model where the spectral density is formed
by zero-width discrete states.  Indeed, each time a new decay channel
opens $d\Gamma_{H_Q} /d m_Q$ experiences a jump ($\Gamma_{H_Q}$
is continuous) , so that
immediately above threshold $d\Gamma_{H_Q} /d m_Q$ is larger than
the smooth OPE curve, in the middle between two successive thresholds
it crosses the smooth prediction, and immediately below the next
threshold $d\Gamma_{H_Q} /d m_Q$ is lower than the OPE-based
expectation. 

The amplitude of oscillations can be estimated as follows.  
Let us present the total width $\Gamma_{H_Q}$ as 
\begin{equation}
\label{g-g}
\Gamma_{H_Q} =\sum_{n=0}^{n=\infty}\Gamma_n -\sum_{M_n>M_{H_Q}} \Gamma_n
\;.
\end{equation}
Widths $\Gamma_n$ are exclusive widths 
of two-body decays, $H_Q\to
\phi + h_n$, where $h_n$ is the $n$-th excited $q\bar{q}_{sp}$ state
with the mass $M_n$. For $M_n>M_{H_Q}$ widths $\Gamma_n$ are not,
of course, physical ones for $H_Q$ decays but they are well defined. We have
used this presentation in Sec.~\ref{sub:matching} where it was shown that 
the first
term can be related to the wave function of 
$H_Q$ state on one hand and to the matrix element of ${\bar Q}Q $ operator 
on the other.  It is clear that the first term in Eq.~(\ref{g-g}) is 
a smooth function of
$m_Q$ and contains no nonanalytic terms we are going after, they are in 
the second one. 

Thus we need to to know $ \Gamma_n$  for $M_n$ in the vicinity of $M_{H_Q}$.
For $M_n \gg M_{H_Q}$ we found in Sec.~\ref{sub:sumrules} that
$\Gamma_n \propto M_n^{-6}$, see Eq.~(\ref{largen}). To extrapolate to 
the vicinity of $ M_{H_Q}$ we account for the threshold factor,
$\Gamma_n \propto (M_{H_Q}^2 - M_n^2)/M_n^{8}$. The coefficient in this
dependence can be fixed by duality of the second term in Eq.~(\ref{g-g})
to four-fermion operators ${\cal O}_{4q}$ , discussed in 
Sec.~\ref{sub:matching},
\begin{equation}
  \label{mdl}
  - \sum _{M_n> M_{H_Q}} \Gamma _n = c_{4q} 
\frac{\langle H_Q| {\cal O}_{4q}|H_Q \rangle} {2 M_{H_Q}}\;.   
\end{equation}
We did not calculate coefficients $c_{4q}$ but we found that up to numerical 
factor 
\begin{equation}
c_{4q}\sim  \frac{G^2}{2\pi} \, \frac{ \beta^4}{m_Q^4}
\end{equation}
Assuming $m_{sp},\,m_q \lesssim \beta$ we estimate 
$\langle H_Q| {\cal O}_{4q}|H_Q \rangle /2 M_{H_Q}\sim \beta$.
Then our model for $\Gamma_n$ in the range $M_n \gtrsim M_{H_Q}$ is
\begin{equation}
\label{phi+n}
\Gamma_n = 3\pi G^2  \,\beta^7 \,\frac{M_{H_Q}^2 - M_n^2}{M_n^8}\;.
\end{equation}
The sum in in Eq.~(\ref{mdl}) was approximated by the integral using
$M_n^2 = \pi^2 \beta^2\, n$.

Now we can evaluate the sum~(\ref{mdl}) with the better accuracy accounting for
small nonanalytical terms.  The result is 
\begin{eqnarray}
\label{md2}
- \sum _{M_n> M_{H_Q}} \Gamma _n & = &\Delta \Gamma^{\mbox{\tiny OPE}} +
\Delta \Gamma^{\mbox{\tiny osc}}\;, \nonumber\\
\Delta \Gamma^{\mbox{\tiny OPE}}& = &
\frac{G^2\beta}{2\pi}\left[\left(\frac{\beta}{M_{H_Q}}\right)^4 +
\frac{\pi^4}{2}\left(\frac{\beta}{M_{H_Q}}\right)^8\right]\;, \nonumber\\
\Delta \Gamma^{\mbox{\tiny osc}} & = &\frac{3}{2} \pi^3
G^2\beta\left(\frac{\beta}{M_{H_Q}}\right)^8\,\left[x(1-x)-\frac{1}{6}\right]
\end{eqnarray}
where
\begin{equation}
x=\mbox{Fractional Part
of\/} \left(\frac{M_{H_Q}^2}{\pi^2\beta^2}\right)\,,~~~~~x \in [0,1)\;.
\end{equation}

\begin{figure}[h]
\centerline{\epsfbox{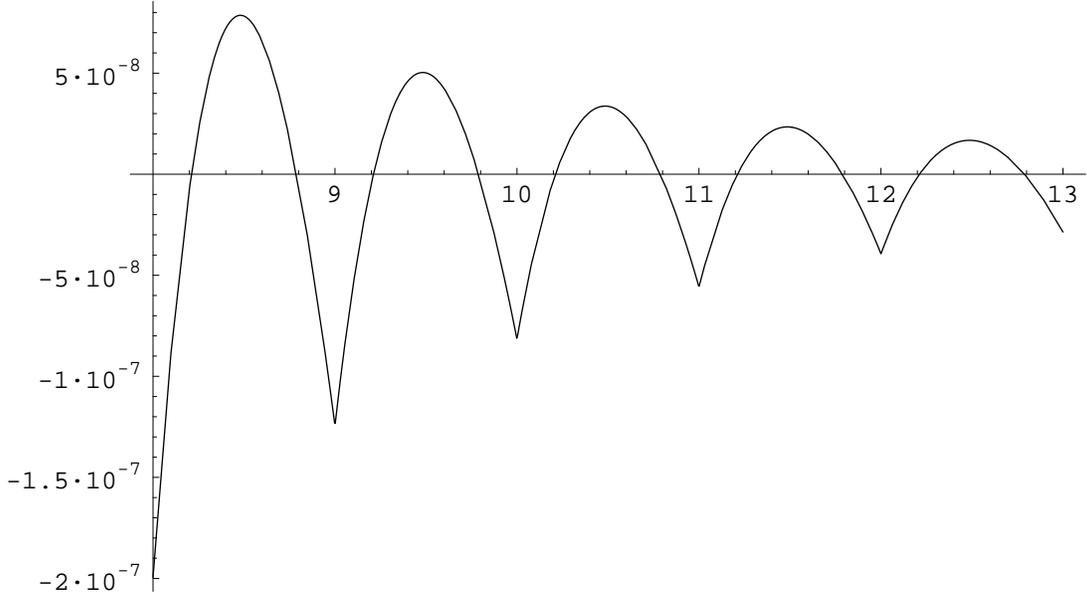}}
\caption{Oscillations in $\Gamma_{H_Q}$. The ratio $\Delta
\Gamma^{\mbox{\tiny osc}}/G^2 \beta$ is  presented as a
function of  $M_{H_Q}^2/\pi^2\beta^2$. }
\label{local}
\end{figure}

The smooth part $\Delta \Gamma^{\mbox{\tiny OPE}}$ in Eq.~(\ref{md2})  is given
by
same OPE term of Eq.~(\ref{mdl}) we discussed above plus higher in 
the power of $1/m_Q$ corrections .
The nonanalytic in
$M_{H_Q}^2$ part $\Delta \Gamma^{\mbox{\tiny osc}}$
 oscillates with period $\pi^2 \beta^2$, see its plot in Fig.~\ref{local}. 
\noindent The amplitude of oscillation is
\begin{equation}
\label{oscil}
\left|\frac{\Delta \Gamma^{\mbox{\tiny osc}}}{\Gamma_Q}\right|_{\rm max}
\sim
\frac{3\pi^4}{2}\,\left(\frac{\beta}{M_{H_Q}}\right)^9\;.
\end{equation}
Note that the derivative $d (\Delta\Gamma^{\mbox{\tiny osc}})/d m_Q$ contains
discontinuities at thresholds, the amplitude of oscillations is larger for
the derivative, 
\begin{equation}
  \label{deroscil}
\left|\frac{d (\Delta \Gamma)^{\mbox{\tiny osc}}/d m_Q}{
d\Gamma_Q/dm_Q}\right|_{\rm max}
\sim
12\pi^2\,\left(\frac{\beta}{M_{H_Q}}\right)^7\;.  
\end{equation}
 
The oscillations under discussion cannot be
produced by any truncated OPE series, they are not seen in the OPE.
Thus the estimate~(\ref{oscil}) gives the actual scale of the expected
duality violations in the problem at hand. Of course, this estimate 
is obtained within a specific model for $\Gamma_n$. The gross features of 
the result are independent of the model, however. They are determined only 
by the fact that the resonances have zero widths. 

It should be noted that the limit $N_c\to \infty$ presents a scenario
which maximizes  duality violations.  In this limit the thresholds
open ``abruptly'', right at the position of the resonances, since the
resonance widths vanish. In the real world of finite $N_c$ the highly
excited states have finite widths, and this effect, on its own, smears
the hadron-saturated cross sections dynamically. If in the zero width
approximation the oscillating duality violating component is
suppressed only by powers of a large parameter ($1/m_Q$ in the case at
hand), switching on finite widths will further suppress the
oscillating component exponentially, see Sec.~\ref{sub:tau} and Ref.~\cite{7}
for further details.

\subsection{Lessons}
\label{sub:lessons}
From the considerations above we conclude that:
\begin{description}
\item[(i)~~] Oscillating terms violating local duality are definitely
  present in the total decay rate (considered as a function of $m_Q$);
\item[(ii)~] Amplitude of oscillations is $\sim {\cal O}(1/m_Q^9)$,
  i.e.  strongly suppressed;
\item[(iii)] If we could average over $m_Q$ in a sufficiently large
  interval the power suppression of the oscillations would turn into
  exponential suppression (in actual QCD, with $N_c=3$, the finite
  resonance widths do a similar job).  Then it is perfectly legitimate
  to consider the OPE-based predictions beyond $1/m_Q^9$.
\end{description}

With this understanding in mind we now turn to the discussion of
duality violations in actual four-dimensional QCD.

\subsection{$\tau$ decays in 1+3 dimensions}
\label{sub:tau}
Let us discuss a quantity of practical interest in 1+3 dimensions
along similar lines, namely the normalized hadronic
$\tau$ width $R_{\tau}$:
\begin{equation}
R_{\tau} \equiv
\frac{\Gamma ( \tau ^- \to \nu _{\tau}+ {\rm hadrons})}
{\Gamma ( \tau ^- \to \nu _{\tau}e^- \bar \nu _e)}
\end{equation}
It can be expressed in terms of spectral densities
$\rho _V$ and $\rho _A$ in the vector and axial-vector
channels, respectively,
\begin{equation}
R_{\tau} = \int _0^{M_{\tau}^2} \frac{{\rm d}s}{M_{\tau}^2}
\left( 1-\frac{s}{M_{\tau}^2}\right) ^2
\left( 1+2\frac{s}{M_{\tau}^2}\right)
\left[ \rho _V(s) + \rho _A(s)
\right] 
=\frac{I_0(M_\tau^2)}{M_\tau^2} -3\,
\frac{I_2(M_\tau^2)}{M_\tau^6}+ 2\,\frac{I_3(M_\tau^2)}{M_\tau^8} \,,
\label{72}
\end{equation}
where the moments $I_n$ are defined as 
\begin{equation}
I_n(M)\;=\; \int_0^{M^2}\; {\rm d}s\, s^n\: \left[\rho_V(s)+\rho_A(s)\right]
\;.
\label{In}
\end{equation}
While $\tau$ decays represent a simpler dynamical problem
than the weak decays of heavy flavor hadrons we have to
simplify it further still before we can arrive at some
definite conclusions. To estimate the oscillating
contribution to $R_{\tau}$ which constitutes duality violation
that cannot be seen in a truncated OPE we consider the
limiting cases of $M_{\tau}$ and $N_c$ large. We will show
that, adopting the resonance model motivated by two-dimensional QCD,
for $N_c \to \infty$ and $M_{\tau}$ large, yet finite,
the duality violation in $R_{\tau}$ scales as
$1/M_{\tau}^6$; for $M_{\tau}\to \infty$ with $N_c$ large,
though finite, the oscillating term is suppressed
exponentially.

Our consideration will be admittedly illustrative. One should not
take literally the numbers we will obtain for many reasons: first of
all the $\tau$ mass is not much larger than the spacing between the
resonances, second, $N_c$ is not large enough to warrant the zero
width approximation. Still we believe that the consideration is
instructive in a qualitative aspect.

For large $N_c$ the spectrum of 1+3 QCD is expected to consist of an
infinite comb of narrow resonances -- in complete analogy to the
't~Hooft model \cite{largenc}.
To keep the closest parallel to it we further assume that
the high excitations in a given channel
(e.g. the vector channel) are equally spaced in $m^2$.
This agrees with the general expectation of a string-like realization
of confinement leading to asymptotically linear Regge trajectories.
The masses of the excited states in, say, the $\rho$ channel are then
given\footnote {In other, less QCD-friendly scenarios, one obtains
instead $m^2_n = m_{\rho}^2 + n/\alpha ^{\prime}$.  The
distinctions between these two scenarios are irrelevant for our
discussion.} by $m^2_n = m_{\rho}^2 + 2n/\alpha ^{\prime}$
\cite{Kprim}, with $\alpha ^{\prime}$ being the slope of the Regge
trajectory (for a review see \cite{KV}). Experimentally one finds
$2/\alpha ^{\prime} \simeq 2~$GeV$^2$.  For large values of $s$ the
spectral densities for both the vector and axial-vector channels will
approach the form:
\begin{equation}
\rho_{\,V} (s) = \rho_A (s) = N_c \cdot \sum_{n=1}^\infty
\delta \left(\frac{s}{\sigma^2}
- n\right)\,;\qquad \sigma^2=\frac{2}{\alpha^{\prime}}\;,
\label{SPECTDENS}
\end{equation}
where a special notation $\sigma^2$ is introduced for $2/\alpha^{\prime}$.
Equation~(\ref{SPECTDENS}) is clearly not expected to hold
at moderate and small values of $s$ where the
vector and axial-vector channels are drastically
different and the resonances are not equidistant.
However, details of the spectral densities at small $s$ play no role in
duality violation.

The contribution of any particular resonance of mass $M_k$ to $R_\tau$,
according to Eq.~(\ref{72}), is given by a simple polynomial in
$1/M_\tau^2$ (times the step function $\theta(M_\tau^2-M_k^2)\,$).
Therefore, variations of properties of resonances below a certain fixed mass
$\mu$  change only the regular terms of the $1/M_\tau^2$ expansion, but have 
no impact on the oscillatory component. From Eq.~(\ref{72}) it is clear that
such variations change only coefficients of the $1/M_\tau^2$, $1/M_\tau^6$ and
$1/M_\tau^8$ terms. It will be clear in what follows that the formal OPE for
$R_\tau$ exactly reproduces these three expansion coefficients as well.

The spectral density in Eq.~(\ref{SPECTDENS}) is dual to the
parton model result; i.e., it coincides with it after
averaging over energy,
\begin{equation}
\langle \rho_{\,V,A} (s)\rangle = N_c\, .
\end{equation}
Thus, the asymptotic  prediction for $R_\tau$ at $M_\tau^2 \to \infty$ is
\begin{equation}
\label{tau0}
R_\tau^0=N_c\;.
\end{equation}

\begin{figure}[h]
\centerline{\epsfbox{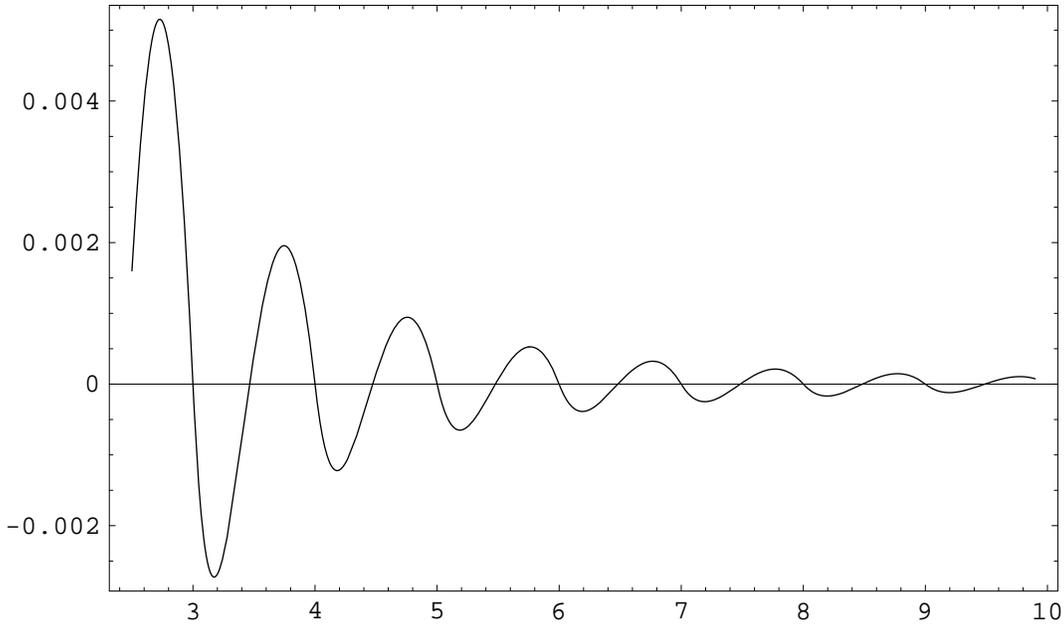}}
\caption{Oscillations in $R_\tau$. The plot of 
$\delta R^{\rm osc}/R_\tau^0$  is presented as 
a function of  $M_\tau^2/\sigma^2$.}
\label{oscillation}
\end{figure}

The sum over resonances in $R_\tau$ is easily calculated analytically:
for the spectral density of Eq.~(\ref{SPECTDENS}) it is
\begin{eqnarray}
R_{\tau} &\,=\,& R_{\tau}^{\mbox{\tiny OPE}} +\delta R^{\rm osc}\;, \nonumber\\
\frac{R_\tau^{\mbox{\tiny OPE}}}{N_c} &\,=\,& 1-\frac{\sigma^2}{M_\tau^2} +
\frac{1}{30}\left(\frac{\sigma^2}{M_\tau^2} \right)^4\;, \nonumber\\
\frac{\delta R^{\rm osc}}{N_c}&\;=\;& -\,
x(1-x)(1-2x)\, \left(\frac{\sigma^2}{M_\tau^2}\right)^3 \,+\,
\left[ x^2(1-x)^2-\frac{1}{30}\right]\,\left(\frac{\sigma^2}{M_\tau^2}\right)^4,
\label{n9}
\end{eqnarray}
where 
$$
x={\rm Fractional~Part~of}\left(\frac{M_\tau^2}{\sigma^2}\right), \;\;\;\: x\in 
[0,1)
\;.
$$
We presented the result as a sum of two functions of $M_\tau^2$, the first 
one, $R_{\tau}^{\mbox{\tiny OPE}}$, is a smooth function expandable in
$1/M_\tau^2$.   The second one,
$\delta R^{\rm osc}$, oscillates with the period $\sigma^2$; its average 
vanishes, see the plot of $\delta R^{\rm osc}/R_\tau^0$ in
Fig.~\ref{oscillation}.

Let us show now that $R_{\tau}^{\mbox{\tiny OPE}}$ coincides with the OPE 
prediction in the model. Power corrections can be presented in the following 
way:
\begin{equation}
R_\tau^{\mbox{\tiny OPE}}\;=\; N_c\,+\, \frac{\tilde I_0}{M_\tau^2}
\,-\, 3\frac{\tilde I_2}{M_\tau^6}
\,+\, 2\frac{\tilde I_3}{M_\tau^8}\;,
\label{n1}
\end{equation}
where the ``condensates'' $\tilde I_n$ are
\begin{equation}
\tilde I_n\;=\; 
\int_0^{\infty}\; {\rm d}s\, s^n\: \left[\rho_V(s)+\rho_A(s)-2N_c \right]
\;.
\label{tildIn}
\end{equation}
This integral representations for the ``condensates'' $\tilde I_n$ follows from
Eqs.~(\ref{72}),~ (\ref{In}) if one assumes that the spectral densities approach 
their 
asymptotic limits faster than any power of $1/s$. In the model at hand, with
 the comb-like spectral density,  the integral
representation (\ref{tildIn}) requires regularization.
As a regularization one can introduce the weight factor $\exp(-\epsilon s)$, 
taking
the limit  $\epsilon \to 0$ at the end. With this regularization, 
$R_\tau^{\mbox{\tiny OPE}}$ from Eq.~(\ref{n9}) is reproduced.

To justify the procedure it is
 instructive to make one
step back and follow more literally the original OPE procedure. Namely,
our primary object of interest is the Euclidean polarization operator
\begin{equation}
\Pi(Q^2)\;=\; \frac{1}{\pi}\,\int  {\rm d}s \:
\frac{\rho_V(s)+\rho_A(s)}{s+Q^2}\;.
\label{n2}
\end{equation}
The original meaning of $\tilde I_n$ is the coefficients in the
asymptotic expansion of $\Pi(Q^2)$ at $Q^2 \to \infty\,$:
\begin{equation}
\pi \Pi(Q^2)\;=\; -2N_c \ln{Q^2} \,+\, {\rm const} \,+\,
\sum_{n=0}^{\infty}\; (-1)^n \frac{\tilde I_n}{\left(Q^2\right)^{n+1}}
\;.
\label{n3}
\end{equation}

The comb of model of Eq.~(\ref{SPECTDENS}) was considered in
Ref.~\cite{Dike},
\begin{equation}
\pi \Pi(Q^2)\;=\; 2N_c\, \sum_{k=1}^{\infty}
\frac{1}{( Q^2/\sigma^2) +n}\:+\:{\rm const}\;=\;
-2N_c \left[\psi\left(\frac{ Q^2}{\sigma^2}\right) \,+\,
\frac{\sigma^2}{Q^2}\right] \:+\:{\rm const}
\;,
\label{n4}
\end{equation}
where $\psi$ is Euler's $\psi$ function.
The asymptotic expansion of $\Pi(Q^2)$ in the model takes the form
\begin{equation}
\pi \Pi(Q^2)\;=\;
-2N_c \left[\ln\frac{Q^2}{\sigma^2} \,+\,
\frac{1}{2} \frac{\sigma^2}{ Q^2} \,-\, \sum_{n=1}^{\infty} \,
\frac{B_{2n}}{2n}\,
\left(\frac{\sigma^2}{ Q^2}\right)^{2n}\right]
\:+\:{\rm const}
\;,
\label{n5}
\end{equation}
where $B_n$ stand for the Bernoulli numbers.  Taking the coefficients of 
$1/Q^2$, $1/Q^6$ and $1/Q^8$ terms from this equation we find the consistency 
with
$R_\tau^{\mbox{\tiny OPE}}$ of Eq.~(\ref{n9}). (The term $1/Q^6$ is absent and 
$1/Q^8$ is defined by $B_4=-1/30$.)
 
Note that the spectral density (\ref{SPECTDENS}) is not literally
``QCD-compatible'': the corresponding $\Pi(Q^2)$ has the $1/Q^2$ nonperturbative
correction forbidden in QCD \cite{2}. This can be easily cured by
adding the resonance with $n=0$ with a half weight, which just
would amount to adding $N_c \sigma^2/Q^2$ to $\pi
\Pi(Q^2)$.  Since this obviously does not change $\delta R^{\rm osc}$ at any 
value of
$M_\tau$, this is inessential for us.

Let us discuss now the duality-violating $\delta R^{\rm osc}(M^2_\tau)$ 
(see Fig.~\ref{oscillation}).
Its dominant component  scales as $1/M_\tau^6$. It is intriguing to
note that the very same scaling law was obtained in Ref.~\cite{Dikeman}
from totally different considerations invoking instantons.
This oscillating component vanishes at $M_\tau$ corresponding to the
new thresholds, and at one point in the middle between the successive
resonances; only the second derivative of $R_\tau$ has a jump at the
thresholds,
\begin{equation}
\left.\frac{1}{R_\tau} \left(\frac{{\rm d}}{{\rm d}M_\tau^2}\right)^2
R_\tau\right|^{x=0}_{x=1}\;=\; 6\frac{\sigma^2}{M_\tau^6}\;.
\label{jump}
\end{equation}
This is the consequence of the threshold factor
$(1-s/M_\tau^2)^2$ in Eq.~(\ref{72}). One power in it is just the
two-body phase space factor $|\vec{p}\,|/M_\tau$. In $1+1$ dimensions
one would have instead $1/|\vec{p}\,|$, the threshold singularity, and
the duality-violating component would be enhanced correspondingly.

The average of $\delta R^{\rm osc}$ vanishes while the amplitude of oscillations
amounts to
\begin{equation}
\left| \frac{\delta R^{\rm osc}}{R_\tau}\right|_{\rm max}\;=\;
\frac{1}{3\sqrt{12}}\, \left(\frac{\sigma^2}{ M_\tau^2}\right)^3
\;.
\label{n12}
\end{equation}
It is interesting that, in spite of the fact that
$M_\tau^6 \,\delta R^{\rm osc}$ is given by a polynomial between the
thresholds, the whole function is well approximated by
\begin{equation}
-\frac{1}{3\sqrt{12}}\, \sin{\left(2\pi\frac{M_\tau^2}{\sigma^2}\right)}
\, \left(\frac{\sigma^2}{ M_\tau^2}\right)^3
\;.
\label{n13}
\end{equation}
Note that the numerical coefficient in Eq.~(\ref{n13}) is rather small, compare 
e.g. to
Eq.~(\ref{jump}).  This suppression is related to the fact that the 
characteristic
scale is $\sigma^2/2\pi$ rather then $\sigma^2$.

Taking our estimate of the oscillation amplitude at its face value
and  using the actual value of the 
$\tau$ mass in Eq.~(\ref{n12}) we find $\delta R^{\rm osc}/R_\tau\sim 3\%$.
It is clear that this is a very
crude estimate given the fact that in the actual $\tau$ decay 
$\sigma^2/M^2_\tau
\sim 2/3$ and we deal with one oscillation at most. 

In the real world with $N_c =3$ we expect a further suppression of
deviations from duality due to the nonvanishing widths of the
resonances naturally smearing out the amplitude of the oscillation. A
rough estimate of this effect can be given in close analogy to
Ref.~\cite{7}. Let us introduce a dimensionless constant $B$
representing the width-to-mass ratio:
\begin{equation}
\frac{\Gamma _n}{m_n} = \frac{B}{N_c} ( 1 + {\cal O}(1/N_c))
\; ;
\end{equation}
i.e., $B$ stays finite for large $N_c$. One actually
guestimates $B \sim 0.5$. Then we infer (see \cite{7} for details)
\begin{equation}
\frac{\Delta R_{\tau}}{R_{\tau}^0} =
\frac{1}{3\sqrt{12}}
\left( \frac{\sigma^2}{M_{\tau}^2}\right) ^3
{\exp}\left(
- \frac{2\pi B M^2_{\tau}}{N_c\sigma^2}
\right)\;.
\end{equation}
The power-suppressed oscillations eventually turn into exponentially
suppressed, although at a larger energy.

\section{Nonvanishing  $\lowercase{m}_\psi$ and comments on the literature}
\label{sec:nonvanishing}
The work~\cite{5} stimulated our interest in the 't~Hooft model as a
laboratory for exploring heavy quark expansions in inclusive decays,
and the implementation of duality.  The authors of Ref.~\cite{5}
compared the decay width of a heavy flavor hadron in the parton
approximation with the result obtained by summing over all exclusive
transition rates for $H_Q \to h_i h_j$.  A systematic excess of the
total width $\Gamma_{H_Q}$ over its parton value $\Gamma_{Q}$ was
observed and was fitted to be
\begin{equation}
\label{deviat}
\frac{\Gamma_{H_Q}-\Gamma_Q}{\Gamma_Q} \sim
\frac{0.15}{m_Q}\;.
\end{equation}
The authors interpreted this access as a violation of duality.
 According
to our understanding (see Sec.~\ref{sub:global}) it should rather be called
breaking of OPE.

Our analytical treatment does not support such a conclusion.  In
particular, we have proved the absence of linear in $1/m_Q$
corrections not only by the OPE method but by the direct analysis of
the 't~Hooft equation (see Sec.~\ref{sec:match}).  Although we tried to closely
follow the analysis of Ref.~\cite{5}, still there is a difference of
kinematical nature. We considered the $\psi$ fields (leptons or
quarks) to be massless, while $m_\psi=m_q \neq 0$ in Ref.~\cite{5}.
The choice of $m_\psi=0$ allows us to limit ourselves to the point
$q^2=0$ where great simplifications occur, in particular, only the massless pion
state  is produced in the $\psi {\bar \psi}$ channel. Analytic
solution of the problem turns out to be possible.

We have checked that the analytical expression for the triple overlap
integral of Ref.~\cite{5} reduces at $q^2=0$ to a simple overlap
integral~(\ref{excl1}) of two wave functions. The simple structure of
(\ref{excl1}) combined with the completeness of the wave functions is
sufficient to prove a perfect match between OPE and hadronic
saturation, at the level of $1/m_Q^4$.

To make full contact between our results and those of Ref.~\cite{5} we
must now consider the impact of $m_\psi \neq 0$. 
Needless to say that at $m_\psi\neq 0$ the parton result for
$\Gamma_Q$ changes,
\begin{equation}
  \label{gamnot}
 \Gamma_Q(m_\psi\neq 0) = \Gamma_Q(m_\psi = 0)\left[1 -
\frac{2m_\psi^2}{m_Q^2} + {\cal O} \left( \frac{m_\psi^4}{m_Q^4}\right)
\right]\;. 
\end{equation}
This effect was certainly included in the analysis
of Ref. \cite{5}. It is also accounted for in OPE as a change of the 
coefficient of the operator ${\bar Q} Q$.

The leading effect due to $m_\psi \neq 0$ is linear in $m_\psi$,
however. Thus, one can wonder whether it produces  $1/m_Q$ corrections.
In the next subsection we will show that the linear in $m_\psi$ 
corrections to the total width are suppressed as $1/m_Q^3$.  
Moreover, 
this is a leading effect which 
produces a distinction between semileptonic and nonleptonic total widths,
all other effects which differentiate them are suppressed as $1/m_Q^4$.
It is clear then that the $1/m_Q$ violation of duality claimed in Ref.~\cite{5}
if it would be present in the nonleptonic width must have been present in
the semileptonic width as well. 

Corrections $1/m_Q^3$ come also from quadratic in $m_\psi$ terms in OPE.
As it was mentioned above these corrections do not differentiate 
semileptonic and nonleptonic widths. We estimate them in 
Sec.~\ref{sub:correction2}. We also estimate in Sec.~\ref{sub:local}
effects of novanishing $m_\psi$ for violations of local duality.
Overall, we conclude that the OPE approach shows that the nonvanishing 
 $m_\psi$ results in $1/m_Q^3$ corrections which are numerically small 
and cannot explain the alleged deviation (\ref{deviat}).
 
On the hadronic saturation side we have checked that the triple
overlap integral of Ref.~\cite{5} is expandable in $m_\psi$, and the
leading correction is quadratic in $m_{\psi,q}$.  Assuming that
$m_\psi\lesssim \beta$ and performing an {\it analytic} summation of the
widths using the expressions of Ref.~\cite{5} for the amplitudes in
conjunction with the sum rules for the hadronic polarization tensor
${\rm Im}\, \Pi(q^2)$ similar to those in
Eqs.~(\ref{SR1},\ref{SR2},\ref{SR3}), we obtained exactly the same
$m_{\psi}^2/m_Q^2$ correction as in Eq.~(\ref{gamnot}). This holds for
the inclusive width smeared over a small interval of masses as
discussed in Sec.~\ref{sub:local}.  Moreover, we checked the exact matching
with OPE at
the level $1/m_Q^3$.

\subsection{Linear in $m_\psi $ corrections}
\label{sub:correction1}
All effects due to $m_\psi \neq 0$ reside 
in $\Pi_{\mu\nu}$, see Eq.~(\ref{Pi1}). The linear in $m_\psi$ part 
in OPE for $\Pi_{\mu\nu}$ is
\begin{equation}
  \label{linm}
  \Delta\Pi_{\mu\nu}(q)= \frac{4m_\psi}{q^2}\,
\langle 0|{\bar \psi}\psi|0\rangle
\left(\frac{q_\mu q_\nu}{q^2} - g_{\mu\nu}\right)
\;.
\end{equation}
It is easy to see that this effect is nothing but a shift of the pion 
($\phi= [\psi {\bar \psi}]_0$) mass from zero,
\begin{equation}
  \label{pionmass}
  \mu_\phi^2 =-\frac{4\pi}{N_c} m_\psi \langle 0|{\bar \psi}\psi|0\rangle
\;. 
\end{equation}
Indeed, comparing Eq.~(\ref{piq}) and Eq.~(\ref{linm}) we see that the
latter is the first order term in $\mu_\phi^2$ expansion of the pion 
pole $1/(q^2- \mu_\phi^2)$ in $\Pi_{\mu\nu}=-(1/\pi)(q_\mu q_\nu-q^2
g_{\mu\nu})/(q^2- \mu_\phi^2)$. Thus, in order to take into account linear in
$m_\psi$ effects in the transition operator all one needs to do is to calculate the
decay $Q \to \phi + q$ with the nonvanishing pion mass.  The result is
 \begin{equation}
\label{psicor}
\frac{\Delta \Gamma_{H_Q}^{(1)}}{ \Gamma_Q} =  \mu_\phi^2 
\frac{2m_Q m_q}{(m_Q^2 -m_q^2)^2}=-\frac{8\pi}{N_c}
m_\psi\langle 0|{\bar \psi}\psi|0\rangle
\frac{m_Q m_q}{(m_Q^2 -m_q^2)^2}
\;.
\end{equation}
At large $m_Q$ it falls off as $1/m_Q^3$. An extra suppression $m_q/m_Q$
is specific for two dimensions.

As for for the numerical value of the correction~(\ref{psicor}) we 
we will substitute $\langle\bar{\psi}\psi \rangle =-N_c \beta/\sqrt{12}$ 
in the chiral limit~\cite{6,Burkardt1}.  For the
value of $m_\psi=m_q=0.56 \beta$ adopted in Ref.~\cite{5} we get
\begin{equation}
\frac{\Delta \Gamma_{H_Q}^{(1)}}{\Gamma_Q} =
2.27 \left( \frac{\beta}{m_Q}\right)^3\;.
\end{equation}

\subsection{Quadratic in $m_\psi $ corrections} 
\label{sub:correction2}
The analysis of the previous subsection refers to the operator 
$\bar{Q}Q\bar{\psi}\psi$ in OPE for the total width. Another operator 
generated 
 due to $m_\psi\neq 0$ is
the four-fermion operator of the type
$\bar{Q}Q\bar{q}q$. It  appears from the
graph of Fig.~\ref{4q0} with lepton lines substituted by $\psi$ lines.
 Calculation of this graph is a simple
exercise. The result for the corresponding correction to the width can
be presented in the following form
\begin{equation}
\label{mcorr}
\Delta \Gamma_{H_Q}^{(2)} = N_c
\frac{G^2 m_\psi^2}{m_Q^2} \, 
\frac{\langle H_Q|\bar{Q}\gamma_5 q\,{\bar q}\gamma_5 Q |H_Q\rangle}{2M_{H_Q}}
\;.
\end{equation}
Up to the factor $N_c$ the same contribution appears in the semileptonic width.
In difference with the operator $\bar{Q}Q\bar{\psi}\psi$ the large $N_c$ limit
does not allow us to factorize 
the  matrix element of Eq.~(\ref{mcorr}). We  will use the factorization 
to get an estimate,
\begin{equation}
\langle H_Q|\bar{Q}\gamma_5 q\, {\bar q}\gamma_5 Q |H_Q\rangle \sim 
-\frac{1}{2N_c} \langle 0| {\bar q} q|0\rangle \,
\langle H_Q|\bar{Q}Q|H_Q\rangle\;.
\end{equation}
As in the previous subsection we use the chiral limit value 
$\langle \bar{q}q \rangle =-N_c \beta/\sqrt{12}$ of the 
vev~\cite{6,Burkardt1} and 
$m_\psi=m_q=0.56 \beta$ for the numerical estimate,
\begin{equation}
\frac{\Delta \Gamma_{H_Q}^{(2)}}{\Gamma_Q} \sim
0.57 \left( \frac{\beta}{m_Q}\right)^3\;.
\end{equation}

\subsection{Local duality violations at $m_\psi \neq 0$}
\label{sub:local}
Here we discuss the impact of new thresholds opening in the spectral
density of the vector $\psi\gamma_\mu\psi$ currents, as $m_Q$
increases.  At $m_\psi=0$ the duality in the $\langle {\bar\psi}
\gamma_\mu \psi, {\bar\psi} \gamma_\nu \psi \rangle$ correlation
function is perfect, since, due to bosonization, the theoretical
expression for this correlator reduces to exactly one massless state
propagation with a known coupling constant.  If $m_\psi \neq 0$ then
higher $ [\psi{\bar\psi} ]$ mesonic states appear in the imaginary
part of $\langle {\bar\psi} \gamma_\mu \psi, {\bar\psi} \gamma_\nu
\psi \rangle$ with residues proportional to $m_\psi^2$.

To calculate the  dependence of the total width $\Gamma_{H_Q}$ on
$m_Q$ near thresholds, i.e. near
\begin{equation}
\label{thres}
M_{H_Q}= M_n^ {[\psi{\bar\psi} ]} +M_k^{[q{\bar q_{sp}} ]}\;,
\end{equation}
we need to find the exclusive width $\Gamma_{nk} (H_Q \to
[\psi{\bar\psi} ]_n + [q{\bar q_{sp}}]_k)$.  We did it in 
Sec.~\ref{sub:oscillating}
for $m_{\psi}=0$ when only the $n=0$ massless $\phi$ state is produced in
the $[\psi{\bar\psi} ]$ channel.

Now, the highly excited states can appear in this channel. They are
pseudoscalar ones corresponding to even values of $n$ (scalar states are not
produced by the conserved current ${\bar\psi}\gamma^\mu \psi $). 
Moreover, with nonvanishing
$ M_n^ {[\psi{\bar\psi} ]}$ the amplitude of transition to scalar 
$[q{\bar q_{sp}}]$
states (odd $k$) is not proportional to $|\vec p\, |$.  Thus, near the
threshold~(\ref{thres}) the exclusive width of decay into the
 pair of the pseudoscalar $[\psi{\bar\psi} ]$ and the scalar $[q{\bar q_{sp}}]$
is singular
because  the factor
$1/|\vec p\, |$ in the phase space that explode at thresholds.  Therefore,
exactly at threshold,
$\Gamma_{nk}$ is infinite. It goes without saying that it cannot
coincide with the smooth OPE prediction near the threshold. These
spikes are clearly visible on the plots of Ref.~\cite{5}.

To maximize the exclusive width $\Gamma_{nk}$ for the decay
$H_q \to [\psi{\bar\psi} ]_n + [q{\bar q_{sp}}]_k$  near the
threshold~(\ref{thres}) we choose the range 
of $M_n^ {[\psi{\bar\psi} ]}$ and $M_k^{[q{\bar q_{sp}} ]}$ 
close to the corresponding partonic configuration.   In terms of 
partons we deal with the decay $Q \to \psi + {\bar \psi} +q$.
If we fix the square of the effective mass of 
$\psi {\bar\psi}$ pair $M^2_ {\psi{\bar\psi}}$   then the momentum of the light
$q$ is
$(m_Q^2 - M^2_ {\psi{\bar\psi}})/(2 m_Q)$. For hadrons 
$ M_n^ {[\psi{\bar\psi}]}$ close to $M^2_ {\psi{\bar\psi}}$ and the
characteristic hadronic mass squared  in $q{\bar q}_{sp}$ system is
$M^2_{[q{\bar q}_{sp}]}\sim
\beta(m_Q^2 -M^2_ {\psi {\bar \psi}})/m_Q$. Then, it is simple to check 
that the threshold range is 
\begin{equation}
M_{H_Q} - M_n^{[\psi{\bar\psi} ]} \sim \beta\;,\;\;\;\; M_k^{[q{\bar q_{sp}} ]}
\sim \beta\;.
\end{equation}
Thus, it is just a few first scalar excitations in $q{\bar q_{sp}}$ system 
that
are relevant. If $M_n^ {[\psi{\bar\psi} ]}$ and $M_k^{[q{\bar q_{sp}} ]}$
fall outside the kinematics indicated above then the exclusive width 
$\Gamma_{nk}$ is additionally suppressed.

Let us remind that the bulk contribution into the total width is
provided by the ground state in the $\psi{\bar\psi} $ channel and 
$M_k^{[q{\bar q_{sp}} ]}\lesssim \sqrt{m_Q\beta}$. What is under discussion 
is a
small tail of highly exited $\psi{\bar\psi} $ states which determines the 
oscillating component.

The relevant matrix element is
\begin{equation}
\langle [\psi{\bar\psi} ]_n [q{\bar q_{sp}}]_k|
({\bar q} \gamma_\mu Q) ({\bar\psi}\gamma^\mu \psi) |H_Q
\rangle=
  \langle [\psi{\bar\psi} ]_n |{\bar\psi}\gamma^\mu
\psi|0\rangle \, \langle  [q{\bar q_{sp}}]_k|  {\bar q}
\gamma_\mu Q|H_Q \rangle\, .
\end{equation}
It is easy to see that the residue $\langle [\psi{\bar\psi} ]_n
|{\bar\psi}\gamma^\mu\psi|0\rangle$ scales as $\sqrt{N_c} m_\psi \beta
/M_n^{[\psi {\bar\psi}]}$, while the transition amplitude $\langle
[q{\bar q_{sp}}]_k| {\bar q} \gamma_\mu Q|H_Q \rangle$ scales as
$\sqrt{m_Q\beta}$. Assembling all factors together we find that
near the threshold
\begin{equation}
\Gamma_{nk} (H_Q \to [\psi{\bar\psi} ]_n + [q{\bar q_{sp}}]_k) \sim
G^2 N_c \frac{m_\psi^2 \beta^3}{m_Q^3 |{\vec p}\,|} \sim \Gamma_Q
\,\frac{m_\psi^2
\beta^3}{m_Q^4 |{\vec p}\,|} \, .
\end{equation}
This equation describes both, the singularity at thresholds and 
deviation from local duality between the thresholds. The spatial
momentum $ |{\vec p}\,|$ is
\begin{equation}
|{\vec p}\,| \approx \sqrt {2 M_k^{[q{\bar q}_{sp}]}\,(M_{H_Q} - M_n^{[\psi
{\bar \psi}]} -
M_k^{[q{\bar q}_{sp}]}) }\; .
\end{equation}

In the middle between the thresholds in $M_n^{\psi {\bar \psi}}$ the
value of $|{\vec p}\,|\sim (\beta^3/m_Q)^{1/2}$, and
  \begin{equation}
    \label{nk}
\Gamma_{nk} (H_Q \to [\psi{\bar\psi} ]_n + [q{\bar q_{sp}}]_k) \sim
\Gamma_Q \, \frac{m_\psi^2 \beta^{3/2}}{m_Q^{7/2}}\;.
  \end{equation}
 This estimate gives the amplitude of the oscillating component. It is
  applicable also to the threshold spikes provided these spikes are
  averaged over the intervals of $M_{H_Q}$ less but comparable 
with the period of oscillations in $M_{H_Q}$ dependence ($\sim
\pi^2\beta^2/M_{H_Q})$.

  Therefore, we conclude that the violation of local duality dies off
  as $1/m_Q^{7/2}$.  This effect is by far the largest duality
  violating contribution.  The occurrence of a relatively weak
  suppression is due to (a) the zero resonance width approximation;
  (b) the singular nature of the two-body phase space in two
  dimensions. Both features have no parallel in actual QCD.

  In summary, we identified two leading effects that are responsible
  for deviations from the parton formula -- one is associated with the
  four-fermion operator in OPE, appearing due to $m_\psi\neq 0$, the
  second is the additional duality violating component that was absent
  at $m_\psi =0$.  The first, inclusive one, dies off as $1/m_Q^3$,
  the second (exclusive) at least as $1/m_Q^{7/2}$. We do not see any
  room for $1/m_Q$ deviations, even oscillating.

\section{Discussion and conclusions}
\label{sec:discussions}
The situation we encounter in the 't~Hooft model is very instructive.
The model is readily treatable, which allows one to
advance quite far in constructing the OPE series. It is
superrenormalizable, thus providing an especially clean environment
for testing various subtle aspects of OPE. The perturbative series for
the coefficient functions in the large $N_c$ limit converges.
We find, with satisfaction, that all general statements regarding OPE
are fully confirmed.

The model also clearly exhibits the breaking of local duality by
oscillating terms. These oscillations are related to the exponential
terms in the Euclidian domain and not seen in OPE. Due to zero
meson widths in the large $N_c$ limit they are suppressed  only by
powers of $1/m_Q$ which we have determined.

We note that in the t' Hooft model the local duality of the OPE
predictions in the inclusive heavy flavor decays (both semileptonic
and nonleptonic) holds much better than in $R(e^+e^-)$, the generally
recognized classical laboratory for applications of OPE. It is in
contrast to the opinion often expressed in the literature that OPE is
not applicable in the inclusive heavy quark widths. Moreover, the
numerical computations of Ref.~\cite{5} suggest that OPE width
approaches the (smeared) hadronic ones at a few percent level very
soon, right after a few first channels are open.

In actual QCD, already the first excited states are broad enough and
inconspicuous, leave alone high excitations. When a finite resonance
width is introduced, it immediately leads to dynamical smearing of
the spectral densities, ensuring an exponential suppression of the
oscillating duality violating terms (see Sec.~\ref{sub:tau} and Ref. \cite{7}).
Thus, in terms of actual QCD we are still very far from the solution of
this extremely important problem; the exercise performed
gives us some kind of an upper bound.

As previously, in actual QCD, we have to rely on models while
estimating the exponential/oscillating terms not seen in OPE.  The
choice is not large -- only two models were suggested previously. One
of them is an instanton-based model \cite{Dikeman}, another is a
resonance-based model \cite{7}, close in spirit to estimates we have
presented above.  The instanton-based model is simple and predictive,
but it apparently lacks the sophistication inherent to the phenomenon
in actuality.  In particular, it predicts an oscillating component
$\sim \sin m_Q$, rather than $ \sim \sin m_Q^2$ as would be natural
from the resonance point of view. Thus, the 't~Hooft model teaches us
that the instanton-based estimates cannot be fully true. On the other
hand, the resonance-based model, which works satisfactorily in the
limit of infinitely narrow resonances, does not give a full answer as
to how strongly the oscillating component is suppressed when the
finite resonance widths are switched on.  It is clear that further
steps in developing the existing or engineering new models are needed.

This work presents the first estimate of the duality violations,
from the resonance-related considerations based on a $1/N_c$
expansion, in the practically important problem of the hadronic
$\tau$ decays.  Although not fully conclusive, the results are very
encouraging, and call for expansion of these ideas in other processes.
This is an obvious task for the future.

\acknowledgments
We are grateful to R.~Jaffe, A.~Kaidalov and  M.~Voloshin for useful
discussions.

This work was supported in part by the DOE under the grant number
DE-FG02-94ER40823, by the NSF under the grant number PHY96-05080,
and by the RFFI under the grant number 96-15-96764.


\begin{thebibliography}{99}

\bibitem{1}
K. Wilson, {\it Phys. Rev.} {\bf 179} (1969) 1499; in
Proc. 1971 Int. Symp. on Electron and Photon Interactions
at High Energies, N. Mistry, Ed. (New York, 1972), p.~115;\\
K. Wilson and J. Kogut, {\it Phys. Reports} {\bf 12} (1974) 75;\\
Wilson's ideas were adapted to QCD in
Ref. \cite{2}.

\bibitem{optical}
I. Bigi, M. Shifman, N. Uraltsev and A. Vainshtein,
{\it Phys. Rev.} {\bf D52} (1995) 196.

 \bibitem{Dike}
M. Shifman,
{\em Theory of Preasymptotic Effects in Weak Inclusive Decays},
in Proc. Workshop on {\em Continuous Advances in QCD}, ed. A.
Smilga
(World Scientific, Singapore, 1994), page 249 [hep-ph/9405246];
{\em Recent Progress in the Heavy Quark Theory}, in {\em
 Particles, Strings and Cosmology}, Proc.  PASCOS-95,
ed. J. Bagger {\em et al.} (World Scientific, Singapore, 1996), page 69
[hep-ph/9505289];\\
see also Ref.~\cite{Dikeman}.

\bibitem{2}
M.A. Shifman, A.I. Vainshtein and V.I. Zakharov,
{\it Nucl. Phys.} {\bf B147} (1979) 385; 448;\\
V. Novikov, M. Shifman, A. Vainshtein and V. Zakharov,
{\it Nucl. Phys.}{\bf B249} (1985) 445.

\bibitem{H1}
G. 't~Hooft, {\it Nucl. Phys.} {\bf B75} (1974) 461 [Reprinted in
G. 't~Hooft, {\em Under the Spell of the Gauge Principle} (World
Scientific, Singapore 1994), page 443]; see also Ref. \cite{LTLY}
for a discussion of subtleties in the regularization of the 't~Hooft
model.

\bibitem{H2}
C. Callan, N. Coote, and D. Gross,
{\it Phys. Rev.} {\bf D13} (1976) 1649;\\
M. Einhorn, {\it Phys. Rev.} {\bf D14} (1976) 3451;\\
M. Einhorn, S. Nussinov, and E. Rabinovici,
 {\it Phys. Rev.} {\bf D15} (1977) 2282;\\
I. Bars and M. Green, {\it Phys. Rev.} {\bf D17} (1978) 537.

\bibitem{LTLY}
F. Lenz, M. Thies, S. Levit and K. Yazaki,
{\it Ann.  Phys.} (N.Y.) {\bf 208} (1991) 1.

\bibitem{3}
M. Burkardt and  E. Swanson, {\it Phys. Rev.} {\bf D46} (1992) 5083.

\bibitem{4}
B. Grinstein and P. Mende, {\it Nucl. Phys.} {\bf B425} (1994  )451.

\bibitem{5}
B. Grinstein and R.  Lebed,
{\it Phys. Rev. } {\bf D57} (1998) 1366; [hep-ph/9708396].

\bibitem{6}
A. Zhitnitsky, {\it Phys. Lett.} {\bf B165} (1985) 405;
{\it Phys. Rev.} {\bf D53} (1996) 5821.

\bibitem{7}
B. Blok, M. Shifman, and Da-Xin Zhang, {\it Phys. Rev.}
{\bf D57} (1998) 2691.

\bibitem{Vol}
M. Voloshin, M. Shifman, {\it Yad. Fiz.} {\bf 41} (1985) 187
[{\it Sov. Journ.
Nucl. Phys.} {\bf 41} (1985) 120]; {\it ZhETF} {\bf 91} (1986) 1180
[{\it Sov. Phys. -- JETP} {\bf 64} (1986) 698].

\bibitem{HQET}
E. Eichten and B. Hill, {\it Phys. Lett.} {\bf B234} (1990) 511;\\
H. Georgi, {\it Phys. Lett.} {\bf B240} (1990) 447.

\bibitem{CGG}
J. Chay, H. Georgi and B. Grinstein, {\it Phys. Lett.} {\bf B247} (1990)
399.

\bibitem{BUV}
I. Bigi, N. Uraltsev and A. Vainshtein, {\it Phys. Lett.} {\bf B293}
(1992) 430; (E) {\bf B297} (1993) 477.

\bibitem{NSVZrev}
V. Novikov, M. Shifman, A. Vainshtein, and V. Zakharov,  {\it Fortsch.
Phys.} {\bf 32} (1985) 585.

\bibitem{Burkardt}
M. Burkardt, {\it Phys. Rev.} {\bf D46} (1992) R2751.

\bibitem{Burkardt1}
M. Burkardt,
in Proceedings of {\em Quantum Infrared Physics 1994} 
(Paris, France, 1994) page 438 [hep-ph/9409333].

\bibitem{largenc}
G. 't~Hooft, {\it Nucl. Phys.} {\bf B72} (1974) 461;\\
E. Witten,  {\it Nucl. Phys.} {\bf B149} (1979) 285.

\bibitem{Kprim}
A. Dubin, A. Kaidalov and Yu. A. Simonov,
{\it Phys. Lett.} {\bf B 323} (1994) 41.

\bibitem{KV}
A. Kaidalov, {\it Sov. Phys. Uspekhi}, {\bf 14} (1972) 600;\\
P. Collins, {\it An Introduction to Regge Theory and High Energy
Physics}, (Cambridge Univ. Press, 1977).

\bibitem{Dikeman}
B. Chibisov, R. Dikeman, M. Shifman and N. Uraltsev,
 {\it Int. J. Mod. Phys. } {\bf A12} (1997) 2075.

\end{thebibliography}
\end{document}